\documentclass[letterpape,11pt]{article}
\pdfoutput = 1 
    
\usepackage{physics}    
\usepackage{jheppub} 
\usepackage[utf8]{inputenc}
\usepackage{amsmath}
\usepackage{slashed}
\usepackage{appendix}
\usepackage{caption}
\usepackage{subcaption}
\usepackage{multirow}
\usepackage{float}

\newcommand{\nl}{\nonumber \\[5pt]}

\newcommand{\bmat}[1]{\boldsymbol{#1}}

\newcommand{\lb}{\Big{\lbrack}}
\newcommand{\rb}{\Big{\rbrack}}
\newcommand{\lp}{\Big{(}}
\newcommand{\rp}{\Big{)}}
\newcommand{\lbc}{\Big{\lbrace}}
\newcommand{\rbc}{\Big{\rbrace}}

\newcommand{\nn}{\nonumber}

\newcommand{\be}{\begin{equation}}
\newcommand{\ee}{\end{equation}}
\newcommand{\bea}{\begin{eqnarray}}
\newcommand{\eea}{\end{eqnarray}}
\newcommand{\balign}{\begin{align}}
\newcommand{\ealign}{\end{align}}
\newcommand{\as}{\alpha_s}

\newcommand{\cd}{\cdot}

\newcommand{\bg}{\begin{gather}}
\newcommand{\foma}{\end{gather}}

\newcommand{\noopsort}[1]{}

\newcommand{\vecb}[1]{\mbox{\boldmath $#1$}}
\newcommand{\vecbe}[1]{\mbox{\boldmath ${\scriptstyle #1}$}}

\def\e{\epsilon}

\def\z{\zeta}

\def\<{\langle}
\def\>{\rangle}

\def\g{\gamma}  
\def\d{\delta}

\def\m{\mu}
\def\n{\nu}

\def\z{\zeta}

\def\({\left(}
\def\[{\left[}
\def\){\right)}
\def\]{\right]}

\def\cos{\hbox{cos}}
\def\sin{\hbox{sin}}
\def\ln{\hbox{ln}}

\def \le { \left }
\def \ri { \right}

\newcommand{\ben}{\begin{eqnarray}}
\newcommand{\een}{\end{eqnarray}}

\newcommand{\bef}{\begin{figure}[htb]\centering}
\newcommand{\eef}{\end{figure}}

\title{Gluon Sivers function in dijet production at the EIC}

\author[a,b]{Miguel G. Echevarria,}
\author[c]{Patricia Andrea Gutierrez García,}
\author[c]{ Ignazio Scimemi}
\affiliation[a]{Department of Physics, University of the Basque Country UPV/EHU, 48080 Bilbao, Spain}
\affiliation[b]{EHU Quantum Center, University of the Basque Country UPV/EHU}
\affiliation[c]{Dpto. de F\'{i}sica Te\'{o}rica \& IPARCOS, Universidad Complutense de Madrid, Plaza de Ciencias 1, E-28040 Madrid, Spain}

\emailAdd{miguel.garciae@ehu.eus}
\emailAdd{patricgu@ucm.es}
\emailAdd{ignazios@ucm.es}

\preprint{IPARCOS-UCM-26-004}
\abstract{The transverse-momentum-dependent (TMD) factorization theorem for dijet production in deep-inelastic scattering is used here to make predictions of the gluon Sivers  function. 
We revise the previously studied unpolarized case and develop the formalism for a transversely polarized target. 
We study the impact of TMD evolution in two different schemes and we use the  current extractions of the evolution kernel at N$^3$LO to make predictions for the future Electron-Ion Collider (EIC).
The results strongly depend on the TMD gluon distributions and their evolution kernel. 
At the EIC, and for the considered models, large values of the Sivers asymmetry higher than 5$\%$ are predicted.}

\date{\today}

\begin{document}

\maketitle
%%%%%%%%%%%%%%%%%%%%%%%%%%%%%%%%%%%%%%%%%%%%%%%%%%%%%%%%%%%%%%%%%%%%%%%%%%%%%%%%%%
\section{Introduction}
The information about quark and gluon non-perturbative dynamics available in
transverse momentum differential cross sections is currently expressed  in terms of transverse momentum dependent  distributions (TMDs) and collinear distributions (PDF or fragmentation functions)~\cite{Angeles-Martinez:2015sea}.
Both depend on collinear momentum fractions and only TMD distributions include a dependence on  quark or gluon  transverse momenta.
The TMD distributions are obtained in the framework of TMD factorization (see~\cite{Becher:2010tm,Collins:2011zzd,Echevarria:2012js,Chiu:2012ir,Vladimirov:2021hdn} for a general treatment with different methods).
In this work we will take the case of dijet production in semi-inclusive deep inelastic scattering (SIDIS).
Dijets have been studied in the small-$x$ limit in \cite{Caucal:2021ent,Caucal:2022ulg,Caucal:2023fsf,Caucal:2023nci}.

The TMD factorization of dijet production in SIDIS  was shown in~\cite{delCastillo:2020omr,delCastillo:2021znl} using Soft Collinear Effective Theory (SCET)~\cite{Bauer:2000yr,Bauer:2001yt,Bauer:2002nz,Beneke:2002ph} (see \cite{Becher:2014oda} for an introduction).
 The factorization theorem reported in those works is readily applicable to the case of polarized targets that allow the study of the gluon Sivers function. 
Several technical issues however deserve a special study that we treat in the present work.

Dijet production is highly sensitive to both quark and gluon TMDs and for this reason it is of most interest at future Eletron-Ion Collider (EIC)~\cite{AbdulKhalek:2021gbh}.
At leading power in transverse momentum and leading order in perturbation theory, both quark and gluon channels contribute (while in single jet SIDIS the gluon effects appear just as corrections in both expansions~\cite{Rodini:2023plb,delCastillo:2023rng,Jaarsma:2025ksf,Piloneta:2025jjb}). 
We consider the cases of polarized and unpolarized targets. The former allows access to the T-odd gluon Sivers TMD ($f_{1T}^{\perp g}$). 
This function describes the distribution of unpolarized
gluons in a transversely polarized nucleon and can be essential in other studies of transverse single-spin asymmetries
(see e.g.~\cite{Boer:2015vso}).
A classification of all gluon TMDs can be found in~\cite{Mulders:2000sh}.
The gluon momentum distributions inside a hadron are notoriously difficult to extract from experimental data. The difficulty comes mainly from the fact that they appear together with light quark distributions that often dominate the processes. 
Some early attempts to extract the gluon Sivers functions can be found in
\cite{DAlesio:2017rzj,DAlesio:2018rnv,DAlesio:2020eqo}, which however do not provide a detailed discussion of factorization nor include TMD evolution effects. In general the processes which are more sensitive to gluon distributions include quarkonia production (as in measurements at the LHCspin facility~\cite{LHCspin:2025lvj}), open di-heavy quark, dijet and di-hadron production in 
semi-inclusive deep inelastic scattering (SIDIS)
\cite{Yuan:2008vn,Godbole:2014tha,Boer:2016fqd,Mukherjee:2016qxa,Bacchetta:2018ivt,Zheng:2018ssm,DAlesio:2019qpk}. 
At present, there is no established model for the non-perturbative part of the gluon Sivers function, but there are studies dedicated to this issue e.g.~\cite{Bacchetta:2024fci}.

In general, gluon Sivers effects can be associated with two independent gauge-link structures, often referred to as Weizsäcker-Williams (WW) and dipole gluon Sivers functions. 
The SIDIS dijet process considered here selects the WW-type gluon Sivers function~\cite{Dominguez:2011wm}, consistently with the color flow and the TMD factorization theorem used below.
In the following, unless explicitly stated otherwise, “gluon Sivers function” refers to this WW-type distribution.

The TMD evolution used in this work is defined by the so called $\zeta$-prescription~\cite{Scimemi:2018xaf} and it is provided by the code \texttt{artemide}~\cite{web}.
This has been successfully tested
 in fits of unpolarized distributions in Drell-Yan and SIDIS \cite{Scimemi:2017etj,Bertone:2019nxa,Scimemi:2019cmh,Moos:2023yfa,Moos:2025sal} and polarized quark Sivers distributions~\cite{Bury:2021sue}
including a perturbative TMD evolution kernel up to next-to-next-to-next-to leading order (N$^3$LO).
The availability of (quark and gluon) unpolarized and (quark) Sivers TMD functions extracted from data within a single scheme at high perturbative order results to be advantageous with respect to other extractions that do not offer a complete treatment of both unpolarized and Sivers cases in the same setup (see the extractions of unpolarized functions in ref.~\cite{Bacchetta:2022awv,Khalek:2021gxf,Bacchetta:2024qre}).
 
In the present work 
we observe that we can define two possible schemes to implement TMD evolution. Out of these, one was already proposed in ref.~\cite{delCastillo:2021znl} (we will call it "CCS-scheme", for the product Collinear-soft--Collinear-soft--Soft) and one is proposed here for the first time (we will call it "${\cal M}$-scheme").
The difference comes from implementing the $\zeta$-prescription and renormalization set-up scales in two different ways.
In the CCS-scheme one defines a $\zeta$-prescription condition for soft function appearing in the dijet factorization theorem independently from the renormalization scales of the collinear-soft functions.
In the 
 ${\cal M}$-scheme one defines an ${\cal M}$-function as the product of the collinear-soft and soft functions appearing in the dijet factorization theorem and the $\zeta$-prescription and renormalization scales are defined for the ${\cal M}$-function.
The ${\cal M}$-scheme is here tested for the first time on unpolarized and Sivers cross sections. The results of the two schemes are consistent within errors, but the ${\cal M}$ is technically simpler and avoids several constraints on renormalization scale choices that were found in ref.~\cite{delCastillo:2021znl}.

As a technical aspect,
we have currently two different values of the non-perturbative part of the evolution kernel in the \texttt{artemide} code.
In fact we have two possible extractions, one which used only Drell-Yan data~\cite{Moos:2023yfa} and one which used Drell-Yan plus SIDIS data~\cite{Moos:2025sal}.
The TMD evolution kernel is universal, however for the moment there is some tension at large $b$ between the two extractions that has not been fully understood yet. We check the prediction for the Sivers asymmetry in the two cases.

Concerning the model for gluon Sivers functions we make some controlled assumptions and quantify uncertainties. As default we consider gluon Sivers functions equal to the sea-quark contributions obtained in fits of the Sivers asymmetry \cite{Bury:2021sue} that we comment later in the text. The associated uncertainty is still definitely large, but nevertheless the results that we get provide valuable information.

The work notation is defined in sec.~\ref{sec:notation}.
The expressions for the factorized cross section for SIDIS are in sec.~\ref{sec: factorizedXsec}. Here we give details on both quark and gluon channel contributions to the cross section.
The definition of the operators that appear in the cross section is reported in sec.~\ref{sec:operators}. The evolution differential equations and the formal expression of the cross sections including the evolution kernels are reported in sec.~\ref{sec:kernels}. In section \ref{sec:M-and-W} we provide the definition of the ${\cal M}$-scheme and the final expressions for the hadronic tensors in all schemes.
Sections~\ref{sec:CSkernel} and~\ref{sec:SiversModels} review Collins-Soper kernel and Sivers function models extractions respectively.
The results of this work are discussed in sec.~\ref{sec:results} and we conclude in sec.~\ref{sec:Conclusions}. Some technical formulas are left for the appendix.

\section{Notation and kinematics}
 \label{sec:notation}

\begin{figure}[h]
   \centering
    \includegraphics[width=0.8\linewidth]{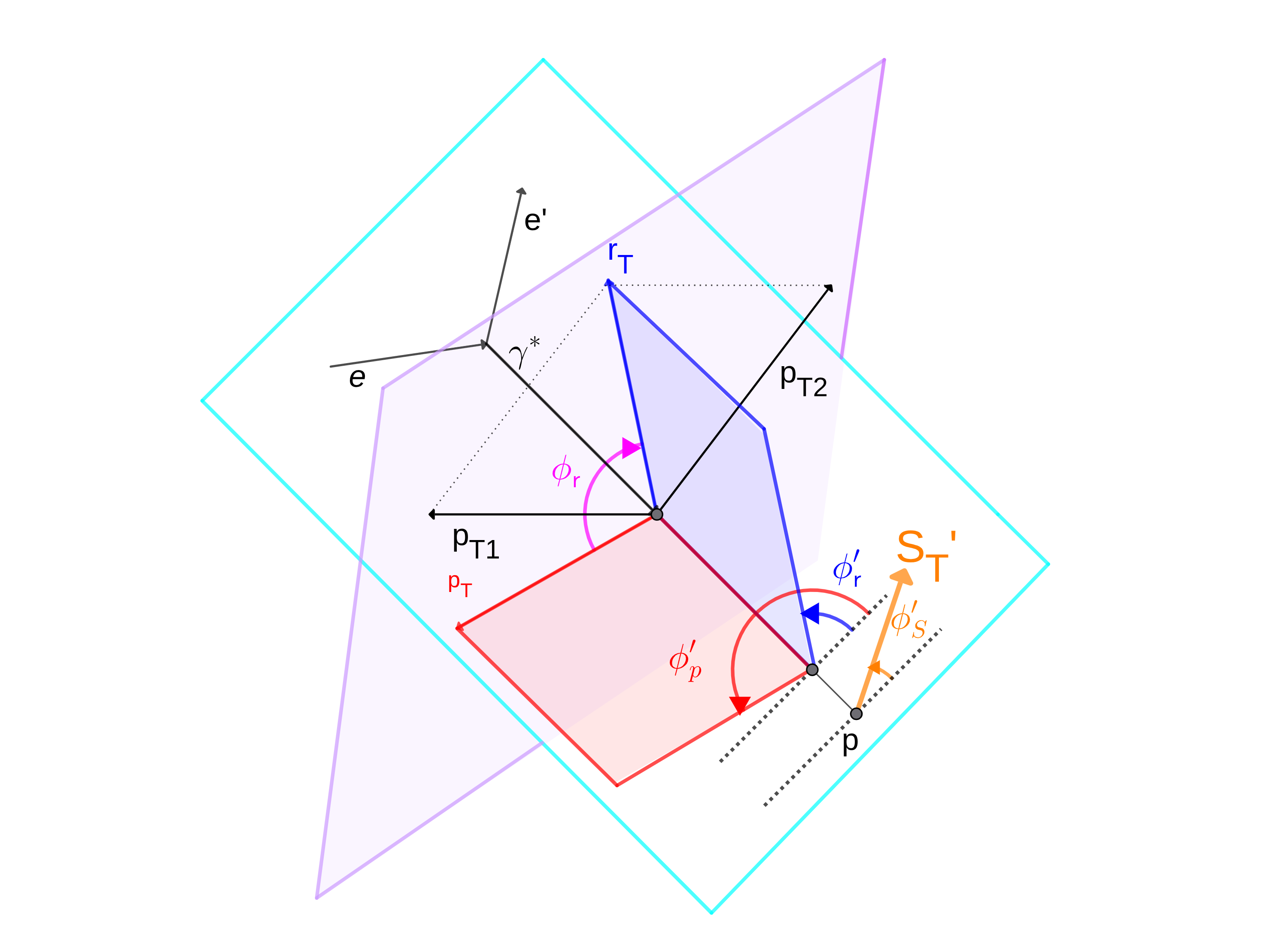}
    \caption{Planes and angles for dijet production in SIDIS in the Breit frame}
    \label{fig:schematic}
\end{figure}

In this work, we study dijet production in hadron-lepton collisions, with the initial hadron being transversely polarized:
\begin{align}\label{eq:process}
\ell + h(P,\uparrow)&\to  \ell'+J_1(p_1) +J_2(p_2) +X.\;
\end{align}
We work in the Breit frame where both the virtual photon and the beam hadron scatter along the z-axis. A schematic view of the process is provided in fig.~\ref{fig:schematic}. 
We use light-cone coordinates to express four-vectors in terms of  their “plus,” “minus,” and transverse components, specifically: 
 \begin{align}
    p^\mu &= p_+ \bar{n}^{\mu} + p_- n^{\mu} + p_{\perp}^{\mu} = (p_+,p_-,\bmat{p}_\perp),
\end{align}
with $n^\mu=(1,0,0,1)/\sqrt{2}$, $\bar n^\mu=(1,0,0,-1)/\sqrt{2}$,
\begin{align}
 p_+ &= n\cdot p,\;\quad  p_- = \bar{n}\cdot p,\;\quad    p^2 = 2p_+ p_- + p^\mu_\perp p_{\mu,\perp}= 2p_+ p_- - \bmat{p}_\perp^2\,,\quad
 p_T=|\bmat{p}_\perp|\,,
\end{align}
where the four-vector $n^\mu$ and $\bar{n}^\mu$ satisfy the relation $n\cdot \bar{n}=1$ with $n^2=\bar n^2=0$. 
The light-cone directions of the two jets are given by $v_1$ and $v_2$. For each of these vector one can define a conjugate one ($\bar v_{1,2}$) such that
\begin{align}
v_J^2 = \bar{ v}_J^2 = 0, \;\quad v_J\cdot \bar{v}_J  = 1,\;\text{ with }  J =1,2
\,, 
\end{align}
The jets  momenta $p_{1,2}^\mu$ have components transverse to the beam axis, $\bmat{p}_{1\perp}$ and $\bmat{p}_{2\perp}$.
 These momenta can be used to define the jet imbalance $\bmat{r}_\perp$ and the hard transverse momentum, $\bmat{p}_\perp$,
\begin{align}
    \bmat{r}_\perp & = \bmat{p}_{1\perp} + \bmat{p}_{2\perp} \,,
    \quad r_T=|\bmat{r}_\perp|\,,
\text{  and  }
    \bmat{p}_\perp  = \frac{\bmat{p}_{1\perp}  - \bmat{p}_{2\perp} }{2},
    \quad
    p_T=|\bmat{p}_\perp|.
\end{align}
At Born level $\bmat{p}_{1\perp} =  - \bmat{p}_{2\perp}$ and thus $ \bmat{r}_\perp  =0 $.  In general the TMD factorization theorem is valid when $r_T\ll p_T$. Following a similar notation for the Fourier  conjugate
variable of $\bmat{r}_\perp$ called $\bmat{b}_\perp$ we can define
$b^\mu=(0,\bmat{b}_\perp,0)$, $b_T=|\bmat{b}_\perp|$.
Similar plane projections are used in~\cite{Kang:2020xez,Chien:2022wiq,Fu:2026nkd}.

%%%%%%%%%%%%%%%%%%%%%%%%%%%%%%%%%%%%%%%%%%%%%%%%%%%%%%%%%%%%%%%%%%%%%%%%%%%%%
Other important variables are the Bjorken momentum fraction ($x$), the total invariant mass squared ($s$) and the inelasticity ($y$) defined as 
\begin{align}
   x&=\frac{Q^2}{2P\cdot q}\,, \qquad s=(\ell+P)^2\,, \qquad y=\frac{q\cdot P}{\ell\cdot P}=\frac{Q^2}{x s}\,,
\end{align}
with $Q^2 = -q^2 = - (\ell - \ell')^2$,  $q^\mu$   the momentum of the virtual photon and $P^\mu$ the momentum of the target hadron. In the Breit frame and neglecting mass corrections we have that the momentum associated to the virtual photon and the target hadron are:
\begin{align}
    q^\mu = (0,0,0,Q),\; \quad P^{\mu}=\frac{1}{2x}(Q,0,0,-Q).
\end{align}
The ratio of the longitudinal momenta of the incoming parton and the target hadron is denoted by $\xi = k^+/P^+$ where $k^{\mu}$ is the momenta of the parton incoming to the hard process.
%%%%%%%%%%%%%%%%%%%%%%%%%%%%%%%%%
Using the pseudorapidities, $\eta_1$ and $\eta_2$, of the outcoming jets, 
we obtain the following relations for the partonic variables,
\begin{align}
    \hat{s} &=(q+k)^2 =  + 4 p_T^2 \cosh^2(\eta_-) \,, \nl
    \hat{t} &=(q-p_2)^2= -4 p_T^2 \cosh(\eta_-) \cosh(\eta_+) \exp(\eta_1)\,, \nl
    \hat{u} &=(q-p_1)^2= -4 p_T^2 \cosh(\eta_-) \cosh(\eta_+) \exp(\eta_2)\,,
    \nonumber\\
    Q^2&=-(\hat{s}+\hat{t}+\hat{u})\,,    =4 p_T^2\cosh^2(\eta_-)\exp(2\eta_+)\,, \nonumber\\
    \xi&=2 x\cosh(\eta_+)\exp(-\eta_+)\,
    \text{  with }
    \label{eq:kindedf}
    \eta_{\pm} = \frac{\eta_1 \pm \eta_2}{2}\,.
\end{align}

The angles in Breit frame shown in fig.~\ref{fig:schematic}, are
\begin{align}
    \phi_r &= \phi_r' - \phi_p', \qquad \phi_S = \phi_S' - \phi_p',
\end{align}
where $\phi_S'$, $\phi_r'$ and $\phi_p'$ are measured w.r.t. the lepton plane and  they are the azimuthal angles of  transverse spin vector of the incoming proton $\bmat{S}_T$, jet imbalance $\bmat{r}_\perp$ and average transverse momenta of the jets $\bmat{p}_\perp$, respectively.
As we will see later in the next section, in order to measure the Sivers function, one needs to look at the $\sin(\phi_S - \phi_r)$ modulated cross section. In the following, we summarize the factorization formalism for dijet production in SIDIS.

%%%%%%%%%%%%%%%%%
\section{Factorization for the dijet cross section in SIDIS}
\label{sec: factorizedXsec}
%%%%%%%%%%%%%%%%%%%%%

The goal of this section is to obtain a factorized cross section for dijet production  in SIDIS when the jet imbalance $\mathbf{r}_\perp$ is small with respect to
the  hard scale transverse momentum of the jets, that is, when the factorization condition is
$r_T\ll p_T$ and the virtual photon and target-hadron directions are back-to-back (Breit frame). For the EIC case this condition is fulfilled for $p_T \in [5, 40]$ GeV and in the central rapidity region as studied in \cite{delCastillo:2021znl}. 
We also assume that there is no hierarchy between the partonic Mandelstam variables, i.e. $\hat{s}\sim |\hat{t}|\sim |\hat{u}|$. Outside this region there are large logarithms in the hard
factor of the cross-section which potentially would ruin the convergence of perturbative expansion. Under the proper factorization conditions, the dijet cross-section is given by
%%%%%%%%%%%%%%%%%%%
\begin{figure}[h]
    \centering
\includegraphics[width=0.70\linewidth]{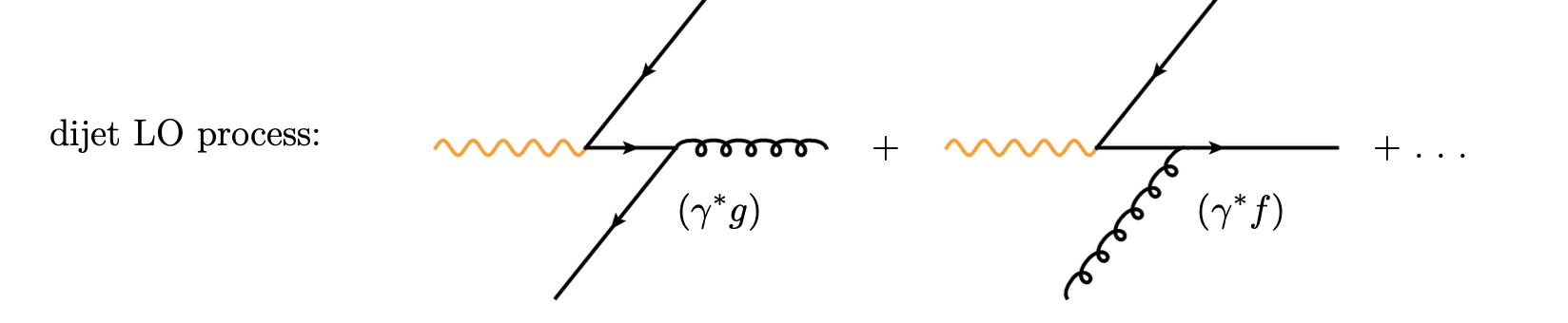}
    \caption{Dijet  hard processes at leading order, from \cite{delCastillo:2020omr}. Here $f=q,\;\bar q$.}
    \label{fig:DHard}
\end{figure}
%%%%%%%%%%%%%%%%%%%%%%%%
\begin{equation}
\label{eq:dsUT}
 \frac{d\sigma}{d\Pi d\bmat{r}_\perp} \equiv
    \frac{d\sigma}{dx d\eta_1 d\eta_2 dp_T d\bmat{r}_\perp} = d \sigma\left(\phi_{S}, \phi_{r}\right)=d \sigma^{U}\left(\phi_{r}\right)+d \sigma^{T}\left(\phi_{S}, \phi_{r}\right),
\end{equation}
where all variables are defined in sec.~\ref{sec:notation} and 
\begin{align}
\label{eq:dPi}
 d\Pi=dx d\eta_1 d\eta_2 dp_T\,.   
\end{align}
In eq.~(\ref{eq:dsUT}),
$d\sigma^U$
corresponds to the case of scattering  from an unpolarized hadron, while  $d\sigma^T $ arises when the hadron is transversely polarized.
We do not treat here the case of longitudinally polarized hadrons, as discussed later in sec.~\ref{sec:gc}.
The formalism for factorization follows~\cite{delCastillo:2020omr} for the unpolarized case, see also fig.~\ref{fig:DHard}.
For each piece of the cross section we have two contributions coming from photon-gluon and photon-quark/anti-quark scattering
\begin{align}
d\sigma^{U,T}&
    =
    d\sigma^{U,T}(\gamma^* g)+\sum_{q=\text{flav.}}d\sigma^{U,T}(\gamma^* q)+\sum_{\bar q=\text{flav.}}d\sigma^{U,T}(\gamma^* \bar q)
\end{align}
and "flav." stands for  quark flavors.
In the following  we describe $(\gamma^* g)$ and $(\gamma^* q)$ channels one by one.

\subsection{Gluon channel}\label{sec:gc}
For the gluon channel the differential cross-section is expressed as \footnote{In these formulas we have not specified the dependence of the functions on the direction. However, there is an implicit dependence on the directions $v_1$ and $v_2$ for the two jet functions and $n$ for the TMDPDF that is explicitly written out later.}
\begin{align}\label{eq:FactFormula}
    \frac{d\sigma (\gamma^* g)}{d\Pi d\bmat{r}_\perp} &=
    \sum_{q} H^{ \mu \nu}_{\gamma^*g\rightarrow q\bar q}(\mu) \int \frac{d^2 \bmat{b}_\perp}{(2\pi)^2}  \, \exp(i \bmat{b}_\perp \cdot \bmat{r}_\perp) \,F_{g, \mu \nu}(\xi, \bmat{b}_\perp, \mu,\zeta_1) \nonumber \\ & 
    \times S_{\gamma g}(\bmat{b}_\perp,\eta_1, \eta_2,\mu,\zeta_2) \, \mathcal{C}_{q}(\bmat{b}_\perp,R,\mu)  J_{q}(p_T,R,\mu)  \mathcal{C}_{\bar{q}}(\bmat{b}_\perp,R,\mu)  J_{\bar{q}}(p_T,R,\mu) \;.
\end{align}
%%%%%%%%%%
%
where $F_{g}^{\mu\nu}$ is the TMD gluon tensor, $H^{\mu\nu}_{\gamma^*g\rightarrow q\bar q}$ is the gluon hard tensor, $S_{\gamma g}$ denotes the dijet soft function and $J_{q}=J_{\bar{q}}$ and $\mathcal{C}_{q}=\mathcal{C}_{\bar{q}}$ are the quark jet and collinear-soft functions.  Notice that the TMD are evaluated at $\xi$ given in sec.~\ref{sec:notation}.
The explicit definition of 
$S_{\gamma g}$, $J_{q(\bar{q})}$, $\mathcal{C}_{q(\bar{q})}$  as matrix element of operators is reported in sec.~\ref{sec:operators}.
Finally, $\mu$ is the factorization scale, and $\zeta_i$'s are the rapidity scales typical of TMD factorization. All scales cancel in the product of all these functions with the condition $\zeta_1\zeta_2=p_T^2$, see~\cite{delCastillo:2020omr}. The variable $\xi$ is the Bjorken collinear momentum fraction and $R$ is the jet radius (for a first estimate we assume to be the same for both jets).

The gluon hard tensor $H^{\mu\nu}(\mu)_{\gamma^*g\rightarrow q\bar q}$ incorporates contributions from both unpolarized ($\propto  H^{U_g}_{\gamma^*g\rightarrow q\bar q}$)
and linearly polarized gluons ($\propto H^{L_g}_{\gamma^*g\rightarrow q\bar q}$) and it does not depend on hadron polarization:
\begin{equation}\label{eq:hadgtens}
    H^{\mu\nu}_{\gamma^* g \to q\bar{q}} = \sigma_0^{g U}\, H^{U}_{\gamma^*g\to q \bar{q} } \frac{g_{\perp}^{\mu\nu}}{d-2} + \sigma_0^{g L} \,H^{L}_{\gamma^*g\to q \bar{q} } \lp -\frac{g_{\perp}^{\mu\nu}}{d-2} +\frac{v^{\mu}_{1\perp}\, v^{\nu}_{2\perp} + v^{\mu}_{2\perp}\, v^{\nu}_{1\perp} }{2 \;v_{1\perp} \cdot v_{2\perp}}\rp 
    \,.
\end{equation}
The hard prefactors are reported in appendix \ref{app:Prefactors}.
  Linearly polarized gluons result to be $\alpha_s^2$ suppressed in the matching with collinear distributions~\cite{Gutierrez-Reyes:2019rug} so that we  neglect them and we use
\begin{equation}\label{eq:hadgtens2}
    H^{\mu\nu}_{\gamma^* g \to q\bar{q}} \simeq \sigma_0^{g U}\, H_{\gamma^*g\to q \bar{q} }^{U} \frac{g_{\perp}^{\mu\nu}}{d-2} 
\end{equation}
The transverse momentum-dependent (TMD) gluon tensor $F_{g}^{\mu\nu}$ can be also decomposed into unpolarized and linearly polarized parts,

\begin{align}\label{eq:decomposition}
F_{g/A}^{\m\n[U]} (\xi,\bmat{r}_\perp) &= 
g_\perp^{\m\n}f_1^g(\xi,r_{T})
+\dots \,,\nn\\
F_{g/A}^{\m\n[T]} (\xi,\bmat{r}_\perp) &=-\
g_\perp^{\m\n} \frac{\e_\perp^{\alpha \beta} r_{\perp\alpha} S_{\perp\beta}}{M} f_{1T}^{\perp g}(\xi,r_{T})
+\dots
\,.
\end{align}
and the dots include TMDs that do not enter this work. see for instance ref.~\cite{Echevarria:2015uaa,Boer:2016xqr}\footnote{In the definition of \cite{Boer:2016xqr}  the gluon distributions are multiplied by a factor $\xi$ ($x$), because of a different normalization of the matrix element definition. We use the  operator definition of sec.~\ref{sec:operators}.}. 
We use $\epsilon_{0123}=-\epsilon^{0123}=1$, and $\epsilon_\perp^{12}=-\epsilon_\perp^{21}=1$ so that $ \epsilon_\perp^{\alpha\beta}r_{\perp\alpha}S_{\perp\beta}=r_T S_T\sin(\phi_S-\phi_r)$.
Using the Fourier transform definition,
\begin{align}
{ G}_{g/A}^{\m\n[pol]} (\xi,\vecb b_\perp) &=
\int d^2\vecb r_\perp\,
e^{-i\vecbe r_\perp\cd \vecbe b_\perp}\,
G_{g/A}^{\m\n[pol]} (\xi,\vecb r_\perp)
\,,
\end{align}
we have
\begin{align}\label{eq:decompositionips}
{ F}_{g/A}^{\m\n[U]} (\xi,\vecb b_{\perp}) &= 
g_\perp^{\m\n}{ f}_1^g(\xi,b_{T})+\dots
\,,\nn\\
{ F}_{g/A}^{\m\n[T]} (\xi,\vecb b_{\perp}) &=
+i g_\perp^{\m\n} \epsilon_{\perp}^{\alpha\beta} M 
b_{\perp\alpha} S_{\perp\beta}
{ f}_{1T}^{\perp g}(\xi,b_{T})
+\dots
\,.
\end{align}
and $ \epsilon_\perp^{\alpha\beta}b_{\perp\alpha}S_{\perp\beta}=b_T S_T\sin(\phi_S-\phi_b)$. The Fourier transforms are
\begin{align}\nn
\label{eq:FT}
    f_1^{g}(\xi,b_T)&= 2\pi \int_0^\infty d r_T r_T J_0(b_T r_T) f_1^{g}(\xi, r_T) \,,\\
    f_{1T}^{\perp g}(\xi,b_T)&= \frac{2\pi }{M^2} \int_0^\infty d r_T  \frac{r_T^2}{b_T}  J_1(b_T r_T) f_{1T}^{\perp g}(\xi, r_T) 
\end{align}
where $J_n$ is the Bessel function of the first kind of order $n$ and it is defined as
\begin{equation*}
    J_n(z)= \frac{1}{2\pi i^n} \int_0^{2\pi} d\varphi \exp(i n \varphi + i z \cos{\varphi})  \,.
\end{equation*}

Separating the cross section for unpolarized and polarized  hadron targets the cross section for the gluon channel is, 
\begin{align}\label{eq:FactFormulaUT}
    \frac{d\sigma (\gamma^* g)}{d\Pi  d\bmat{r}_\perp} &= \frac{d\sigma^U (\gamma^* g)}{d\Pi d\bmat{r}_\perp}+ \frac{d\sigma^T (\gamma^* g)}{d\Pi d\bmat{r}_\perp}\,.
    \end{align}
 Using
\begin{equation}
    F_{g, \mu \nu}^U H^{\mu\nu}_{\gamma^* g \to q\bar{q}} =  \, \sigma_0^{gU} H^U_{\gamma^*g\to q \bar{q} }  f_{1}^{g}(\xi, b)\,,
\end{equation}
the unpolarized part agrees with \cite{Pisano:2013cya,Boer:2016fqd,Efremov:2017iwh,Efremov:2018myn} and it is
\begin{align}
     \frac{d\sigma^U (\gamma^* g)}{d\Pi d\bmat{r}_\perp} &=
    \sum_{q} \sigma_0^{gU}H^U_{\gamma^*g\to q \bar{q} }
     (\mu)\int \frac{d^2 \bmat{b}_\perp}{(2\pi)^2}  \, \exp(i \bmat{b}_\perp \cdot \bmat{r}_\perp) \,f_1^g(\xi, b_T, \mu,\zeta_1) \nonumber\\ & 
    \times S_{\gamma g}(\bmat{b}_\perp,\eta_1, \eta_2,\mu,\zeta_2) \, \mathcal{C}_{q}(\bmat{b}_\perp,R,\mu)  J_{q}(p_T,R,\mu)  \mathcal{C}_{\bar{q}}(\bmat{b}_\perp,R,\mu)  J_{\bar{q}}(p_T,R,\mu)
    \,,
\end{align}
where $ \xi$ is related to $x$ in eq.~(\ref{eq:kindedf}) and we have indicated explicitly the renormalization and rapidity scales $(\mu,\zeta)$.
For the transversely polarized targets  we have
\begin{equation}
    F_{g, \mu \nu}^T H^{\mu\nu}_{\gamma^* g \to q\bar{q}} = +i S_T\, \sigma_0^{gU} H^U_{\gamma^*g\to q \bar{q} }
    \sin (\phi_S - \phi_b) f_{1T}^{\perp g}.
\end{equation}
and the corresponding cross-section part:
\begin{align}
      \frac{d\sigma^T (\gamma^* g)}{d\Pi d\bmat{r}_\perp} &=
     \sum_{q} (+i) \sigma_0^{gU}H^U_{\gamma^*g\to q \bar{q} }
     (\mu)\int \frac{d^2 \bmat{b}_\perp}{(2\pi)^2}  \, \exp(i \bmat{b}_\perp \cdot \bmat{r}_\perp) \,S_T \sin(\phi_S - \phi_b)b_T Mf_{1T}^{\perp g}(\xi, b_T, \mu,\zeta_1) \nonumber\\ & 
    \times S_{\gamma g}(\bmat{b}_\perp,\eta_1, \eta_2,\mu,\zeta_2) \, \mathcal{C}_{q}(\bmat{b}_\perp,R,\mu)  J_{q}(p_T,R,\mu)  \mathcal{C}_{\bar{q}}(\bmat{b}_\perp,R,\mu)  J_{\bar{q}}(p_T,R,\mu).
\end{align}

%%%%%%
\subsection{Quark/anti-quark channels}
\label{subsec:QC}
The differential cross-section for the quark  and anti-quark channels can be obtained one from the other just with the interchange of labels $q\leftrightarrow\bar q$.
 Thus, unless the notation creates confusion, in the following we limit ourselves to  describe the quark case.
 
The quark cross section is
\begin{align}\label{eq:FactGamFU}\nn
    \frac{d\sigma
    (\gamma^* q)}{d\Pi d\bmat{r}_\perp} &= \sigma_0^{qU}\,\sum_{q=\text{flav.}} H^{U}_{\gamma^*q\to g q }(\hat{s},\hat{t}, \hat{u};\mu) \int \frac{d^2\bmat{b}_\perp}{(2\pi)^2}  \, \exp(i \bmat{b}_\perp \cdot \bmat{r}_\perp) \,F_{q}(\xi, \bmat{b}_\perp,  \mu,\zeta_1) \\ &
    \times S_{\gamma q}(\bmat{b}_\perp,\mu,\zeta_2) \mathcal{C}_{g}(\bmat{b}_\perp,R,\mu)  J_{g}(p_T,R,\mu) \mathcal{C}_{q}(\bmat{b}_\perp,R,\mu)  J_{q}(p_T,R,\mu) \,,
\end{align}
where the hard factor $ H^{U}_{\gamma^*q\to g q }$ has the same functional form as in the unpolarized hadron case, and consistent with the absence of polarization in the final-state partons. The hard factor and the dijet soft function are the same for quarks and anti-quarks, i.e.  $S_{\gamma q}=S_{\gamma \bar q}$. The quark TMD parton distribution function $F_{q}(\xi, b)$ is evaluated in $\xi$ as in the the gluon channel case and it  admits a decomposition into its unpolarized and Sivers-function components~\cite{Scimemi:2018mmi},
\begin{align}
\label{eq:qUplusTk}
    F_q(\xi,\bmat{r}_\perp)    &=F_q^{[U]}(\xi,\bmat{r}_\perp)  +  F_q^{[T]}(\xi,\bmat{r}_\perp)\,,\nn \\
    F_q^{[U]}(\xi,\bmat{r}_\perp) &= f_1^q(\xi,r_T) +\dots
    \nn \\
   F_q^{[T]}(\xi,\bmat{r}_\perp) 
 &= 
 -\frac{\epsilon_\perp^{\alpha \beta} r_{\perp\alpha}S_{\perp\beta}}{M} f_{1T}^{\perp q}(\xi,r_T)+\dots,
\end{align}
where we have omitted the functions that are not considered in this work.
The function $ f_1^q$ represents the  distribution of an unpolarized quark in an unpolarized hadron, and $f_{1T}^{\perp q}$, the quark Sivers function, represents the distribution for an unpolarized quark in a transversely polarized hadron.
In conjugate space we have
\begin{align}
\label{eq:qUplusTb}
    F_q^{[U]}(\xi,\bmat{b}_\perp) &=  f_1^q(\xi,b_T) +\dots\,,
    \nn \\
   F_q^{[T]}(\xi,\bmat{b}_\perp) 
    &= +i \epsilon_\perp^{\alpha\beta }b_{\perp\alpha}S_{\perp\beta}  M f_{1T}^{\perp q}(\xi,b_T)+\dots\,.
\end{align}
The Fourier transforms to pass from momentum to conjugate spaces are analogue to eq.~\ref{eq:FT}.
 Using the decomposition  in eq.~(\ref{eq:dsUT}) one gets 
\begin{align}
     \frac{d\sigma^U (\gamma^* q)}{d\Pi d\bmat{r}_\perp}& = 
     \sigma_0^{qU}\,\sum_{q=\text{flav.}} H^{U}_{\gamma^*q\to g q }(\hat{s},\hat{t}, \hat{u};\mu) \int \frac{d^2\bmat{b}_\perp}{(2\pi)^2}  \, \exp(i \bmat{b}_\perp \cdot \bmat{r}_\perp) \,f_1^q(\xi, b_T,  \mu,\zeta_1) \\
     \nonumber
    &\times S_{\gamma q}(\bmat{b}_\perp,\mu,\zeta_2) \mathcal{C}_{g}(\bmat{b}_\perp,R,\mu)  J_{g}(p_T,R,\mu) \mathcal{C}_{q}(\bmat{b}_\perp,R,\mu)  J_{q}(p_T,R,\mu),
\\
    \frac{d\sigma^T (\gamma^* q)}{d\Pi d\bmat{r}_\perp} &= 
  +  i\sigma_0^{qU}\,\sum_{q=\text{flav.}} H^{U}_{\gamma^*q\to g q }(\hat{s},\hat{t}, \hat{u},\mu) \int \frac{d^2\bmat{b}_\perp}{(2\pi)^2}  \, \exp(i \bmat{b}_\perp \cdot \bmat{r}_\perp) \,S_T \sin(\phi_S - \phi_b)b_T M\nonumber \\
    &\times f_{1T}^{\perp q}(\xi, b_T, \mu,\zeta_1) S_{\gamma q}(\bmat{b}_\perp,\mu,\zeta_2) \mathcal{C}_{g}(\bmat{b}_\perp,R,\mu)  J_{g}(p_T,R,\mu) \mathcal{C}_{q}(\bmat{b}_\perp,R,\mu)  J_{q}(p_T,R,\mu).
\end{align}
The symbols used for each functions are the same as in the gluon channel case.

\subsection{Weighted  cross section and Sivers asymmetry 
}

It is useful to introduce a notation for angular integrated cross sections for the unpolarized and polarized cases, namely,
\begin{align}
W^{UU}&=r_T\int_{-\pi}^{\pi} d\phi_r \frac{d\sigma^U}{d\Pi d\bmat{r}_\perp}=\frac{d\sigma^U}{d\Pi d{r}_T}\,,
\nn\\
    W^{UT}&=r_T\int_{-\pi}^{\pi} d\phi_r\sin(\phi_S -\phi_r)  \frac{d\sigma^T}{d\Pi d\bmat{r}_\perp}\,.
 \end{align}
 We   decompose $W^{UU,UT}$ according  to quark and gluon initiated contributions, 
\begin{align}
\label{eq:WUUUT}
    W^{UU,UT} &= W^{UU,UT}_{g} +\sum_{q=\text{flav.}} \left(W^{UU,UT}_{q}+ W^{UU,UT}_{\bar q}\right)\,.
\end{align}
For each channel the factorization formulas are schematically and  respectively given by:
\begin{align}
\label{eq:wuu}
\nn
W^{UU}_{g}&=r_T\int_{-\pi}^{\pi} d\phi_r \frac{d\sigma^U(\gamma^*g)}{d\Pi d\bmat{r}_\perp} 
\\ \nn &
= r_T\sigma_0^{gU}\sum_{q=\text{flav.}} H^{U}_{\gamma^*g \rightarrow q\bar{q}}(\mu)\int\frac{b_T db_T}{2\pi} J_0(r_T b_T) \int_{-\pi}^{\pi}d\phi_b  f_{1}^{ g}(\xi,b) S_{\gamma g} \, \mathcal{C}_{q} \mathcal{C}_{\bar{q}} J_{q}  J_{\bar{q}}\,,
    \nonumber \\
    W^{UU}_{q}&=r_T\int_{-\pi}^{\pi} d\phi_r \frac{d\sigma^U(\gamma^*q)}{d\Pi d\bmat{r}_\perp} = r_T\sigma_0^{qU}\, H^{U}_{\gamma^*q\to g q }(\mu)   \int \frac{b_T db_T}{(2\pi)}  J_0(r_T b_T)   \int_{-\pi}^{\pi} d\phi_b
       f_{1}^{q}(\xi, b_T)
    S_{\gamma q} \mathcal{C}_{g}  \mathcal{C}_{q} J_{g} J_{q} \,,
     \end{align}
     and
     \begin{align}
\label{eq:wut}
\nn
     W^{UT}_{g}&=r_T\int_{-\pi}^{\pi} d\phi_r\sin(\phi_S -\phi_r) \frac{d\sigma^T(\gamma^*g)}{d\Pi d\bmat{r}_\perp}=-r_T \sigma_0^{gU}\sum_{q=\text{flav.}} H^{U}_{\gamma^*g \rightarrow q\bar{q}}(\mu) \\
    \nonumber 
    &\times\int\frac{b_T db_T}{2\pi} J_1(r_T b_T) \int_{-\pi}^{\pi}d\phi_b \sin^2(\phi_S - \phi_b) S_T b_T M f_{1T}^{\perp g}(\xi,b_T) S_{\gamma g} \, \mathcal{C}_{q} \mathcal{C}_{\bar{q}} J_{q}  J_{\bar{q}},
\\
      W^{UT}_q&=r_T\int_{-\pi}^{\pi} d\phi_r \sin(\phi_S - \phi_r) \frac{d\sigma^{T}(\gamma^* q)}{d\Pi d\bmat{r}_\perp}= -r_T \sigma_0^{qU}\, H^{U}_{\gamma^*q\to g q }(\hat{s},\hat{t}, \hat{u},\mu) 
    \nonumber \\
    & \times \int \frac{b_T db_T}{(2\pi)}  J_1(r_T b_T)   \int_{-\pi}^{\pi} d\phi_b\sin^2(\phi_S - \phi_b) 
     S_T b_T M f_{1T}^{\perp q}(\xi, b_T)
    S_{\gamma q} \mathcal{C}_{g}  \mathcal{C}_{q} J_{g} J_{q}  ,  
\end{align}
where we have not reported the argument of all functions for brevity.
In the next section we provide a detailed definition of each matrix element entering these formulas.
In the plots that we show as  our results in sec.~\ref{sec:results} we will  show the partially integrated   cross-sections,
\begin{align}
\label{eq:tutxi}
\nn
     \frac{d\sigma^{UU}}{ dr_T}
  & = \int d{\eta_1}d{\eta_2}\;\delta(\eta_1)  \delta(\eta_2)
  \int dp_T\delta(p_T-p_T^0)
  \int_{x_{\rm min}}^{x_{\rm max}} dx\; W^{UU}\,, 
  \\ \nn
  \frac{\langle d \sigma^{UT}\rangle}{ dr_T }&\equiv
   \int d{\eta_1}d{\eta_2}\;\delta(\eta_1) \delta(\eta_2)\int d\phi_r \;r_T \sin(\phi_S-\phi_r) \int dp_T\delta(p_T-p_T^0) \frac{d\sigma^{UT}}{ d\bmat{r}_T dp_T}\\
&   = \int d{\eta_1}d{\eta_2}\;\delta(\eta_1) \delta(\eta_2)
 \int dp_T\delta(p_T-p_T^0)\int_{x_{\rm min}}^{x_{\rm max}} dx W^{UT}\,, 
\end{align}
and the values of the inputs are specified in sec.~\ref{sec:results}.
The Sivers asymmetries  in sec.~\ref{sec:results} are then defined as the ratio

\begin{equation}
    A^{\text{Sivers}}_{[x_{\text{min}},x_{\text{max}}]}= \frac{\int_{x_{\rm min}}^{x_{\rm max}}dx
    \int 
    %d\phi_S 
    d\phi_r \sin(\phi_r-\phi_S) (d\sigma(\uparrow) - d\sigma(\downarrow))}{\frac{1}{2}
    \int_{x_{\rm min}}^{x_{\rm max}}dx
    \int
    d\phi_r 
    (d\sigma(\uparrow) + d\sigma(\downarrow))}
     = 2 \frac{\langle d \sigma^{UT}\rangle/ dr_T }{d \sigma^{UU}/ dr_T}
     \,,
\end{equation}
where $d\sigma(\uparrow),\;d\sigma(\downarrow)$
are the spin up and spin down contributions to the cross section according to Trento conventions~\cite{Bacchetta:2004jz}.
%%%%%%%%%%%%%%%%%%%%%%%%
\section{Operator definitions}
\label{sec:operators}
Each of the functions of the previous section is understood as free of rapidity divergences. This ensures that we are expressing it in terms of universal functions. This is achieved in the following way.  The TMD are defined starting from the operator definitions 
\begin{align}
\hat{B}_{q} \left(x, \bmat{b}_\perp\right) &=\frac{1}{2} \sum_{X} \int \frac{d \xi^{+}}{2 \pi} e^{-i x p^{-} \xi^{+}}\left\{T\left[\bar{q}_{i} \tilde{W}_{n}^{T}\right]_{a}\left(\frac{\xi}{2}\right)|X\rangle \gamma_{i j}^{-}\langle X| \bar{T}\left[\tilde{W}_{n}^{T \dagger} q_{j}\right]_{a}\left(-\frac{\xi}{2}\right)\right\}, \nl
\hat{B}_{\bar{q}} \left(x, \bmat{b}_\perp \right) &=\frac{1}{2} \sum_{X} \int \frac{d \xi^{+}}{2 \pi} e^{-i x p^{-} \xi^{+}}\left\{T\left[\tilde{W}_{n}^{T \dagger} q_{j}\right]_{a}\left(\frac{\xi}{2}\right)|X\rangle \gamma_{i j}^{-}\langle X| \bar{T}\left[\bar{q}_{i} \tilde{W}_{n}^{T}\right]_{a}\left(-\frac{\xi}{2}\right)\right\}, \nl
\hat{B}_{g, \mu\nu} \left(x, \bmat{b}_\perp \right) &=\frac{1}{x p^{-}} \sum_{X} \int \frac{d \xi^{+}}{2 \pi} e^{-i x p^{-} \xi^{+}}\left\{T\left[F_{-\mu} \tilde{W}_{n}^{T}\right]_{a}\left(\frac{\xi}{2}\right)|X\rangle\langle X| \bar{T}\left[\tilde{W}_{n}^{T \dagger} F_{-\nu}\right]_{a}\left(-\frac{\xi}{2}\right)\right\}\,,
\end{align}
where $\xi= ( \xi^{+}, 0^-, \bmat{b}_\perp )%_n
$. The repeated color indices $a$ ($a=1, \ldots, N_{c}$ for quarks and $a=1, \ldots, N_{c}^{2}-1$ for gluons) are summed up and we have not introduced any renormalization scale in the arguments for simplicity. The functions written above are un-subtracted bare functions, that is, they still have an overlap with soft or opposite modes and are not renormalized.
Using a $\delta$-regulator~\cite{Chiu:2009yx,GarciaEchevarria:2011rb,Echevarria:2015byo,Echevarria:2016scs} the zero-bin subtracted functions are obtained 
\begin{equation}\label{eq:ZBBeam}
\hat{B}_i^{\rm{subt}}(x,\bmat{b}_\perp,\mu,p^- \delta_+) = \frac{ \hat{B}_i (x,\bmat{b}_\perp, \mu,p^- / \delta^-) }{S_i (\bmat{b}_\perp,\mu, \sqrt{\delta^+ \delta^{-}})}\,, \qquad \text{ with } i=q,\bar q, g
\end{equation}
where $S_i$ ($S_i$ depends just on the quark and gluon SU(3)$_c$ representation, that is $S_q=S_{\bar q}$) is the
back-to-back two-direction soft function   for SIDIS and Drell-Yan processes~\cite{Echevarria:2015byo,Echevarria:2016scs},
\begin{equation}
S_i \left( \bmat{b}_\perp\right)=\frac{\operatorname{Tr}}{N_{c}}\langle 0 |T\left[S_{n i}^{T \dagger} \tilde{S}_{\bar{n} i}^{T}\right]\left(0^{+}, 0^{-}, \bmat{b}_\perp \right) \bar{T}\left[\tilde{S}_{\bar{n} i}^{T \dagger} S_{n i}^{T}\right](0) | 0 \rangle\,,
\qquad \text{ with } i=q,g
\end{equation} 
and $S_{n}^{T}$ and $\tilde{S}_{\bar{n}}^{T}$ are soft Wilson lines as defined in \cite{Echevarria:2016scs} for  fundamental and adjoint color representations.

The TMD factorization~\cite{GarciaEchevarria:2011rb,Echevarria:2015byo} is realized  using  the soft function property 
\begin{equation}
S_i  (\bmat{b}_\perp,\mu,\sqrt{\delta^+ \delta^{-}})  = (S_i(\bmat{b}_\perp,\mu,\delta^+  \nu)) ^{\frac{1}{2}}  (S_i(\bmat{b}_\perp,\mu,\delta^-/\nu ))^{\frac{1}{2}}
\,,\qquad \text{ with } i=q,g
\end{equation}
where $\nu$ is an arbitrary positive scale that parametrizes the ambiguity in this splitting and that  will be removed from the final result, introducing this way a constraint on the product of rapidity scales. 
The TMD is finally defined as 
\begin{equation}
\label{eq:TMDdef}
 F_i^\text{bare} (\xi, \bmat{b}_\perp, \mu, \zeta_1 ) = B_i^{\text{subt}} (\xi, \bmat{b}_\perp, \mu, p^-  \delta^+) S_i^{\frac{1}{2}} (\bmat{b}_\perp,\mu,\delta^+ \nu) \Bigg |_{\sqrt{2} \,p^-  \nu \to \sqrt {\zeta_1} } \,,\qquad \text{ with } i=q,\bar q,g
\end{equation}
Similar manipulations can be performed for the other soft functions of the process,
\begin{align}
\label{SF}\nn
   \hat S_{\gamma g}(\bmat{b}_\perp) &= \frac{1}{C_F C_A} \langle 0| S^\dagger_{g n}(\bmat{b}_\perp,-\infty)_{ca'}\text{Tr}\lb S_{qv_2}(+\infty,\bmat{b}_\perp)T^{a'}S^\dagger_{q v_1} (+\infty,\bmat{b}_\perp) \\
    &\times S_{qv_1}(+\infty,0)T^{a}S^{\dagger}_{qv_2} (+\infty,0)\rb S_{gn}(0,-\infty)_{ac}|0 \rangle,
\end{align}
and soft function corresponding to the case of incoming  quark or antiquark  that is obtained exchanging
\begin{equation}
\hat S_{\gamma q} = \hat S_{\gamma g} (n \leftrightarrow v_2)
\,.
\end{equation}
In this  case we have
\begin{equation}
\label{eq:soft-ratio}
S_{\gamma i}^\text{bare} (\bmat{b}_\perp, \mu, \zeta_2)  = \frac{ \hat S_{\gamma i}  (\bmat{b}_\perp,\mu,\sqrt{A_n} \,\delta^+) }{S(\bmat{b}_\perp,\mu,\sqrt{\delta^+ \delta^-})}   S^{\frac{1}{2}} (\bmat{b}_\perp,\mu,\delta^- /\nu)  \Bigg |_{ \nu/ \sqrt{2 A_{n}}  \to \sqrt{\zeta_2}}\,,\qquad \text{ with } i=q,g
\end{equation}
with $A_n=v_1\cdot v_2/[(n\cdot v_1)(n\cdot v_2)]$ and the l.h.s free of rapidity divergences.
In practice  $\hat B$ and $\hat S_{\gamma i}$ can be re-written as a product of TMDs which are free of rapidity divergences,  and we have the factorization condition
\begin{equation}
\label{eq:zz}
\zeta_1 \, \zeta_2 = \frac{(k^-)^2}{A_n} =  \frac{\hat{u} \;\hat{t}}{\hat{s}} =p_T^2\,,
\end{equation}
where $\hat{s}$, $\hat{t}$, and $\hat{u}$ are the partonic Mandelstam variables, eq.~(\ref{eq:kindedf}). The combination $\zeta_1 \,\zeta_2$ is Lorentz invariant.
Notice that in the present case  $\zeta_2$ is a dimensionless quantity but $\zeta_1$, as usual, has dimensions of mass squared. The natural way to choose the values of $\zeta_1$ and $\zeta_2$ is 
\begin{align}
\zeta_1 =p_T^2,\;\quad \zeta_2  =1 .
\end{align}
The other functions that appear in the cross section are defined as in~\cite{delCastillo:2020omr}
\begin{align}
  \mathcal{C}_{q}(\bmat{b}_\perp,R,\mu) &= \int d\bmat{b}_\perp \exp( \bmat{b}_\perp \cdot \bmat{v}\, \bar{v} \cdot \partial) \frac{1}{N_R} \text{Tr} \langle 0|T \lb U_n^{\dagger}W_t (0)\rb \,\Theta_{\text{alg.}} \, \bar{T} \lb W_t^{\dagger} U_n (0) \rb |0\rangle \;,
\nn\\ \nn
 J_{q}(p,R,\mu)&=   \frac{1}{2 \sqrt{2} N_c}\text{Tr}\int d^4 x e^{i p x} \langle 0 |\bar \chi_v(p) \slashed{\bar v}\delta_{\text{alg}}(R) \chi_v(0)|0\rangle\,,\\
\eta_\perp^{\rho\nu} J_{g}(p,R,\mu)&=  - \frac{1}{ (N_c^2-1)}\sum_{A=1}^{N_c^2-1}\int d^4 x (\sqrt{2}\bar v \cdot p) e^{i p x}   \langle 0 |B^{\eta_\perp\rho A}_{v}(x) \delta_{\text{alg}}(R) gB^{\eta_\perp \nu A}_{v}(0)|0\rangle ,
 \end{align}
 where the prefix $\eta_\perp$ here refers to the plane orthogonal to the directions of the jets $v$ and $\bar{v}$ whose  transverse components are obtained with the tensor $\eta_\perp^{\rho\nu} =g^{\rho\nu}-(v^\rho \bar v^\nu+\bar v^\nu v^\rho)/(v\cdot\bar v)$. 
These functions do not suffer from rapidity divergences and have a standard behavior under renormalization. The factor $\delta_{alg}$ refers to the cuts in phase-space that define the jet algorithm. An explicit calculation for the cone and the $k_T$ algorithms can be found in \cite{Kang1001.0014:2012am}. The function $\Theta_{\rm alg}$ ensures that only the contributions from outside the jet will contribute to the jet-imbalance~\cite{delCastillo:2020omr}.

%%%%%%%%%%%%%%%%%%%%
\section{Renormalization group equations and evolution kernels}
\label{sec:kernels}

In this section we sum up the main properties of the anomalous dimensions that appear in the factorization theorem. The anomalous dimension of the dijet soft functions depend on the variable
$c_{\bmat{b}}=\cos\phi_b=\bmat{b}_\perp \cdot \bmat{v}_{1\perp}/(b_T v_{1T})$.
From previous equations and taking into account the rapidity scale dependence one finds that  soft functions  obey
\begin{align}\label{eq:ADSga}
    \frac{d}{d\ln\mu}S_{\gamma i}(\bmat{b}_\perp,\mu,\zeta_2)&=
    \gamma_{S_{\gamma i}}(\mu,c_\mathbf{b})S_{\gamma i}(\bmat{b}_\perp,\mu,\zeta_2)\,, \nn\\
    \frac{d}{d\ln\zeta_2}S_{\gamma i}(\bmat{b}_\perp,\mu,\zeta_2)&=
    -{\cal D}_i(b_T,\mu)S_{\gamma i}(\bmat{b}_\perp,\mu,\zeta_2)\, , \qquad \text{ with } i=q, g
\end{align}
where ${\cal D}$ is  the Collins-Soper kernel  and $(\gamma_{S_{\gamma\bar q}},{\cal D}_{\bar q})=(\gamma_{S_{\gamma q}},{\cal D}_{q})$.
Similar equations hold for the TMD
\begin{align}\label{eq:ADFga}
    \frac{d}{d\ln\mu}F_i(\bmat{b}_\perp,\mu,\zeta_1)&=
    \gamma_{F_i}(\mu)F_i(\bmat{b}_\perp,\mu,\zeta_1)\,, \nn\\
    \frac{d}{d\ln\zeta_1}F_i(\bmat{b}_\perp,\mu,\zeta_1)&=
    -{\cal D}_i(b_T,\mu)F_i(\bmat{b}_\perp,\mu,\zeta_1)\, ,\qquad \text{ with } i=q,\bar q,g\,
\end{align}
with $\gamma_{F_q}=\gamma_{F_{\bar q}}$.
Although apparently similar, eq.~(\ref{eq:ADFga}) is independent of the cosine $c_{\bmat{b}}$ contrary to the dijet soft function case, eq.~(\ref{eq:ADSga}).
The integration of the angle $\phi_b$ is non trivial and in principle can give rise to spurious complex terms in the resummation.
Concerning the other  function that appear in the factorization theorem we have 
\begin{align}\nn
    \frac{d}{d\ln\mu} H_{\gamma i}& =\gamma_{H_{\gamma i}}(\mu)H_{\gamma i},\\ \nn
    \frac{d}{d\ln\mu} {\cal C}_{ i}& =\gamma_{{\cal C} i}(\mu,c_{\mathbf{b}},R){\cal C}_{ i},\\ 
     \frac{d}{d\ln\mu} J_{ i}& =\gamma_{J i}(\mu,R) J_{ i}\, \qquad \text{ with } i=q,\bar q,g\,
\end{align}
where we have made explicit also the dependence on the jet radius $R$ and $c_{\mathbf{b}}$.
Consistency of the cross section requires (here $\gamma_\alpha$ is from the $a_s$ in the hard factor~\cite{delCastillo:2020omr}),
\begin{align}\nn\label{eq:gammasum}
    \gamma_{H_{\gamma g}}(\mu)+\gamma_\alpha(\mu)+\gamma_{F_{g}}(\mu)+
    \gamma_{J_{q}}(\mu,R)+\gamma_{J_{\bar q}}(\mu,R)+
    \gamma_{{\cal C}_{q}}(\mu,c_{\mathbf{b}}, R)+\gamma_{{\cal C}_{\bar q}}(\mu,c_{\mathbf{b}}, R)+\gamma_{S_{\gamma g}}(\mu,c_{\mathbf{b}})&=0,\\
  \gamma_{H_{\gamma q}}(\mu)+\gamma_\alpha(\mu)+\gamma_{F_{q}}(\mu)+
    \gamma_{J_{q}}(\mu,R)+\gamma_{J_{g}}(\mu,R)+
    \gamma_{{\cal C}_{q}}(\mu,c_{\mathbf{b}}, R)+\gamma_{{\cal C}_{g}}(\mu,c_{\mathbf{b}}, R)+\gamma_{S_{\gamma q}}(\mu,c_{\mathbf{b}})&=0.
\end{align}
The one loop values of these anomalous dimensions can be found in the appendix~\ref{app:AD}.
Looking at this equations one can see that if the dijet soft and collinear-soft functions are evolved from the same initial to the same final scale the $c_{\mathbf{b}}$ dependence in the anomalous dimensions cancels altogether making the evolution simpler and well defined.
Nevertheless one should still check the convergence of the perturbative series when such a choice is made.
We discuss more this point in the next sections.

The evolution kernels that are obtained solving the renormalization group equations of each functions can be grouped together using factors $\mathcal{R}_{q,g}$ in the $W^{UU,UT}$ obtained in sec.~\ref{sec: factorizedXsec}.

For the unpolarized cross sections we have 
\begin{align}
    \label{eq:sifg1}
    W^{UU}_g&= r_T\sum_{q=\text{flav.}} \sigma_0^{g}H^U_{\gamma^*g\to q \bar{q} }(\mu)|_{\mu=p_T} (J_{q}(p_T,R,\mu_J) J_{\bar q}(p_T,R,\mu_J)) \nonumber\\
    &\times \int_0^{+\infty} b db\, J_0(b r_T) f_1^g(\xi, b_T) \mathcal{R}_g \lp (\{\mu_k\},\zeta_{1,0},\zeta_{2,0}) \to (p_T,p_T^2,1) \rp  \hat \sigma^{U}_g (b_T, R, \{\mu_i\}),
    \\ \label{eq:siff1}
    W^{UU}_q
  &= r_T \sigma_0^{q}H^{U}_{\gamma^*q\to g q }(\mu)|_{\mu=p_T}( J_{q}(p_T,R,\mu_J) J_{g}(p_T,R,\mu_J) )\nonumber\\ 
    &\times \int_0^{+\infty} b db\, J_0(b r_T) f_1^q(\xi, b_T) \mathcal{R}_q \lp (\{\mu_k\},\zeta_{1,0},\zeta_{2,0}) \to (p_T,p_T^2,1) \rp  \hat\sigma^{U}_q(b_T, R, \{\mu_i\}),
\end{align}
and $\hat\sigma^{U}_{q,g}$ indicate the $\phi_b$-integrated product of collinear-soft  and soft functions, see eq.~(\ref{eq:wuu}). Similar equations can be written for the polarized case with trivial changes in the labels and $\phi_b$ integrations for TMD distributions and the other functions, see eq.~(\ref{eq:wut}). The evolution kernels $\mathcal{R}_{q,g}$ in eq.~(\ref{eq:sifg1}-\ref{eq:siff1}) are the same in the polarized and unpolarized cases.
They can be written as
\begin{align} 
\nonumber 
    \mathcal{R}_g \lp (\{\mu_k\},\zeta_{1,0}^g,\zeta_{2,0}^{S_{\gamma g}}) \to (p_T,p_T^2,1) \rp &= \mathcal{R}_{J_q}(\mu_J \to p_T)^2 \mathcal{R}_{\mathcal{C}_q}(\mu_{\mathcal{C}_q }\to p_T)^2 \\ & \label{eq:r1}
    \times \mathcal{R}_F^g \lp (\mu_0,\zeta_{1,0}^g) \to (p_T,p_T^2) \rp \mathcal{R}_{S}^{g} \lp (\mu_0,\zeta_{2,0}^{S_{\gamma g}}) \to (p_T,1) \rp, \\ \nonumber
\mathcal{R}_q \lp (\{\mu_k\},\zeta_{1,0}^q,\zeta_{2,0}^{S_{\gamma q}}) \to (p_T,p_T^2,1) \rp &= \mathcal{R}_{J_q}(\mu_J \to p_T)
\mathcal{R}_{J_g}(\mu_J \to p_T)\mathcal{R}_{\mathcal{C}_q}(\mu_{\mathcal{C}_q} \to p_T) \mathcal{R}_{\mathcal{C}_g}(\mu_{\mathcal{C}_g} \to p_T)\\ &
    \times \mathcal{R}_F^q \lp (\mu_0,\zeta_{1,0}^q) \to (p_T,p_T^2) \rp \mathcal{R}_{S}^{q} \lp (\mu_0,\zeta_{2,0}^{S_{\gamma q}}) \to (p_T,1) \rp, \label{eq:r2}
\end{align}
where $k=\mathcal{C}_{q,g},\;S_{\gamma q, \,\gamma g},J_{q,\,g}$ and  we have the following notation fo each evolution kernel: $\mathcal{R}_{J_{q,g}}$ is the one for the jet function, $\mathcal{R}_{\mathcal{C}_{q,g}}$ is the one of collinear-soft functions, $\mathcal{R}_F^{q,g}$ the one of TMD and finally $\mathcal{R}_S^{q, g}$ is the one of the dijet soft function.
 Notice that in principle each  evolution kernel is defined at  a scale which minimizes the logs. 
The kernels for single-scale evolution have a standard form and a review up to NLL is given in \cite{Hornig:2016js},
\begin{align}
    \mathcal{R}_{i}(\mu_i \to p_T)&=e^{K_i(\mu_i \to p_T)} \lp \frac{\mu_i}{m_i} \rp^{\omega_i(\mu_i \to p_T)},\qquad i = \{ \mathcal{C}_q, \mathcal{C}_g, J_q, J_g \}
\end{align}
where
\begin{align}
\label{eq:omegaFNLL}
 \omega_i(\mu_i \to p_T) \Big\vert_{\rm NLL}  &=-\frac{\Gamma_{i}^0}{\beta_0} \left[\ln{r}+\left(\frac{\Gamma_{1}}{\Gamma_{0}}-\frac{\beta_1}{\beta_0}\right)\frac{\alpha_s(\mu_i)}{4\pi}(r-1)\right] \,,\\
  \label{KFNLL}
K_i(\mu_i \to p_T) \Big\vert_{\rm NLL} &=-\frac{\gamma_{i}^0}{2\beta_0}\ln {r} - \frac{2\pi\Gamma_{i}^0}{(\beta_0)^2}\bigg[\frac{r-1-r\ln{r}}{\alpha_s(p_T)} \nonumber \\
  &   +\left(\frac{\Gamma_{1}}{\Gamma_{0}}-\frac{\beta_1}{\beta_0}\right)\frac{1-r+\ln{r}}{4\pi}+\frac{\beta_1}{8\pi\beta_0}\ln^2{r}\bigg],
\end{align}
with $r=\alpha_s(p_T)/\alpha_s(\mu_i)$ and the rest of anomalous dimensions  can be found in \cite{delCastillo:2021znl}.
The evolution kernels for TMD, collinear-soft and soft functions resum Sudakov double logs and deserve a special discussion that we provide in the next section.

%%%%%%%%%%%%%%%%%%%%%
\section{Evolution kernel for the soft and collinear-soft function:  ${\cal M}$-scheme
}
\label{sec:M-and-W}
%%%%%%%%%%%%%%%%%%%%%%%%%%
The resummation of the rapidity logarithms is performed here using $\zeta$-prescription \cite{Scimemi:2018xaf}.
The general form of the  kernel for the functions that  have both UV and rapidity evolution is
\begin{align}\label{eq:evolkernel}
\mathcal{ R}\big(\{\m_i,\z_i\}\rightarrow\{\m_f,\z_f\}\big) &=
\le( \frac{\z_f}{\z_{_i}} \ri)^{-\mathcal{D}\le(b_T;\m_i\ri)}\exp\le\{
\int_{\m_i}^{\m_f} \frac{d\bar\m}{\bar\m}\, 
\g\le(\as(\bar\m),\z_f\ri)
\ri\}
\,,
 \end{align}
where $\gamma$ indicated the UV anomalous dimension.
In the case of the $\zeta$-prescription the TMD kernel  is
\begin{equation}
    \mathcal{R}_F^{q,g}(\{\mu_0,\zeta_{0}^{q,g}(b_T)\}\rightarrow\{p_T,p_T^2\}) = \lp \frac{p_T^2}{\zeta_{\mu_f}^{q,g}(b_T) } \rp^{-\mathcal{D}_{q,g}(b_T, p_T)},
\end{equation}
with $\zeta_{\mu_f}^{q,g}$ derived from the  saddle point condition of the null-evolution curve, see  next section and ref~\cite{Scimemi:2018xaf}.
In ref.~\cite{delCastillo:2021znl} the soft functions ($S_{\gamma q, \gamma g}$) evolution kernels  have a formally similar result
\begin{equation}
    \mathcal{R}_S^{q,g}(\{\mu_0^S,\zeta_{2,0}^{S_{\gamma q,\gamma g}}(b_T)\}\rightarrow\{p_T,1\}) = {\cal R}^{c_{\bmat{b}}}_{S,(q,g)}\lp \frac{1}{\zeta_{2,\mu_f}^{S_{\gamma q,\gamma g}}(b_T) } \rp^{-\mathcal{D}_{q,g}(b_T, p_T)},
\end{equation}
and their dependence  on $c_{\bmat{b}}$  makes this direct implementation problematic.
Notably, we recall that the perturbative  $c_{\bmat{b} }$
dependence in the anomalous dimensions is canceled among the soft function and the collinear-soft functions, see eq.~(\ref{eq:gammasum}). 
This cancellation is no more ensured when functions are resummed naively and spurious imaginary terms are generated.
The presence of $c_{\bmat{b}}$ dependent terms appears because of 
the separation of soft and collinear-soft functions and it is possibly related to a passage from SCET-II factorization to a different one, which should be further investigated.  However the presence of imaginary part in the anomalous dimensions may  suggest that the separation  between soft and collinear-soft functions although formally correct is not  actually working.
A solution for the resummation was provided in \cite{delCastillo:2021znl}.
In that work all the terms that contained $c_{\bmat{b}}$ where treated perturbatively and not resummed.  At the same time the resummation involved only the part of the functions that do not depend on this angle.

In this section we start defining a different scheme that we call ${\cal M}$-scheme.
 The idea is to  define the ${\cal M}$-functions
 \begin{align}
    \mathcal{M}_g(\bmat{b}_\perp,R,\mu,\zeta_2) &= \mathcal{C}_{\bar{q}}(\bmat{b}_\perp, R,\mu) \mathcal{C}_{q}(\bmat{b}_\perp, R,\mu) S_{\gamma g}(\bmat{b}_\perp,\mu,\zeta_2) ,\nn\\
      \mathcal{M}_q(\bmat{b}_\perp,R, \mu,\zeta_2)&= \mathcal{C}_{g}(\bmat{b}_\perp, R,\mu)\mathcal{C}_{q}(\bmat{b}_\perp, R,\mu)S_{\gamma q}(\bmat{b}_\perp,\mu,\zeta_2) .
\end{align}
and $\mathcal{C}_{\bar{q}}=\mathcal{C}_{q}$.
The respective  evolution equations are 
\begin{align}
    \frac{d}{d\ln \mu}\mathcal{M}_i&=\gamma_{\mathcal{M}_i}\mathcal{M}_i,\nn\\
      \frac{d}{d\ln \zeta}\mathcal{M}_i&=-{\cal{D}}_i \mathcal{M}_i,
\end{align}
with ($i=q,\bar q,g$)
and  $\gamma_{\mathcal{M}_i}$ independent of $c_b$,
\begin{align}
\gamma_{\mathcal{M}_g}(\mu)&= 2 \gamma_{\mathcal{C}_{q}} (\mu,c_{\mathbf b})+ \gamma_{S_{\gamma g}}(\mu,c_{\mathbf b}), \nn\\
\gamma_{\mathcal{M}_q}(\mu)&= 
       \gamma_{\mathcal{C}_{g}}(\mu,c_{\mathbf b}) +\gamma_{\mathcal{C}_{q}} (\mu,c_{\mathbf b})+ \gamma_{S_{\gamma q}} (\mu,c_{\mathbf b}).
\end{align}
Using the perturbative expansion of the anomalous dimensions
\begin{equation}
\label{eq:gMexpanded}
    \gamma = \sum_{n=1} a_s^n \gamma^{[n]},\qquad  {\cal M} = \sum_{n=1} a_s^n {\cal M}^{[n]},\qquad a_s=\frac{\alpha_s}{4\pi},
\end{equation}
the one loop expression of $\gamma_{\mathcal{M}_i}$ are
\begin{align}
  \gamma_{\mathcal{M}_g}^{[1]}&= - 4 C_A\ln\zeta_2 -8C_F \ln\frac{\hat{s}}{ p_T^2 R^2} \,,\nn\\
    \gamma_{\mathcal{M}_q}^{[1]} 
    & = -4  (C_F + C_A)\ln\frac{\hat{s}}{ p_T^2 R^2}   +  4(C_F-C_A) \ln \lp \frac{\hat{t} }{ \hat{u}  }  \rp   - 4C_F \ln\zeta_2\,.
    \label{eq:GMs}
\end{align}
Notably, this structure exhibits no angular dependence and one can use standard $\zeta$-prescription rules to evolve this function so that one obtains
\begin{align}
{\cal R}_{\mathcal{M}_{q,g}}(\{\mu_0,\zeta_{2,0}^{{\cal M}_{q,g}}(b_T)\}\rightarrow \{p_T,1\})&=\left(\frac{1}{\zeta_{2,\mu_f}^{{\cal M}_{q,g}}(b_T)}\right)^{-{\cal D}_{q,g}(b_T,p_T)} 
\end{align}
as the evolution kernel from the saddle point $(\mu_0, \zeta_{2,0}^{\cal M})$ to the highest scales $(p_T, 1)$.  The perturbative and exact solution of $\zeta_{2,\mu_f}^{\mathcal{M}_{q,g}}$ is reported in appendix \ref{app:Msadp}.

\subsection{Scale fixing and non-perturbative effects}
 The TMD evolution is fundamental for the predictions and the  $\zeta$-prescription, which is already included in the code  \texttt{artemide},  was used to fit the quark Sivers function~\cite{Bury:2020vhj,Bury:2021sue}.
The TMD evolution kernel ${\cal D}$ has been object of many phenomenological researches in refs.~\cite{Moos:2025sal}(ART25),\cite{Moos:2023yfa} (ART23), \cite{Scimemi:2019cmh} (SV19), \cite{Bacchetta:2025ara} (MAPNN), \cite{Bacchetta:2024qre} (MAP24) and \cite{Bacchetta:2022awv} (MAP22);  lattice computations performed in  refs.~\cite{Bollweg:2024zet} (BGMZ24), \cite{Avkhadiev:2023poz, Avkhadiev:2024mgd} (ASWZ24), \cite{Shu:2023cot} (SSSV23), \cite{LatticePartonLPC:2022eev} (LPC22); parton branching  within the CASCADE generator \cite{CASCADE:2021bxe, BermudezMartinez:2020tys} determined by the method of ref.~\cite{BermudezMartinez:2022ctj};   instanton vacuum model \cite{Liu:2024sqj}, thrust distributions in single inclusive electron-positron annihilation data in ref.~\cite{Boglione:2023duo}, and hadron-structure oriented approach \cite{Aslan:2024nqg}.

In the cross section  of the dijet system we have functions with  single  ($\mu$) and double scale ($\mu,\zeta$) evolution . For practical implementation we need to discuss the proper scale at which each function is initially defined.

The simplest object to treat is  the single scale dependent jet functions $J_{q,\bar q,g}$. For that  we evolve  from
$    \mu_{J} = p_T R\label{eq:muJ}\;
$ to the high scale $p_T$. For our EIC predictions of this work we we use
\begin{align}   p_T=20\ \text{GeV}\,,
   \quad R=0.7\,,
\end{align}

The double scale evolution is instead more complicated and also the usage of $\zeta$-prescription is not unique. We show here two possible example of definition of this prescription which are related to 
the  complication  provided by the $c_{\bmat{b}}$ dependence of some of the anomalous dimensions as discussed in sec.~\ref{sec:kernels}.
 In~\cite{delCastillo:2021znl} the evolution of each function was performed with the initial and final scale choices 
\begin{align}
   {\cal R}_{{\cal C}_{q,g}}:&\;   \mu_{\mathcal{C}_{q}} = \mu_{\mathcal{C}_{g}} =2 e^{-\gamma_E}  \lp \frac{1}{b_T} + \frac{1}{b_{\mathrm{max}}^{\cal C}} \rp
\label{eq:muC}\longrightarrow   p_T \,,\\
 {\cal R}_{{\cal S}_{\gamma q}}, \; {\cal R}_{{\cal S}_{\gamma g}}:&\;  (\mu_0,\zeta_{2,0}^{S_{\gamma q,\gamma g}}(b_T))=\left( \frac{2 e^{-\gamma_E}}{b^*},\zeta_{2,0}^{ S_{\gamma q,\gamma g}}(b_T)\right)\longrightarrow (p_T,\;1) \,,\\
 {\cal R}_{F_{q,g}}:\;&
 \;  (\mu_0,\zeta_{1,0}^{(\gamma q),(\gamma g)}(b_T))=\left( \frac{2 e^{-\gamma_E}}{b^*},\zeta_{1,0}^{(\gamma q),(\gamma g)}(b_T)\right)\longrightarrow (p_T,\;p_T^2) \,,
\end{align}
with 
\begin{align}
\label{eq:b*}
  b^*  = \frac{b_T}{\sqrt{1+b_T^2/b_{\mathrm{max}}^2}},
 \end{align}
and the values of the constants are in tab.~\ref{tab1}\footnote{In principle one can choose a different $b_\text{max}$  or $\beta_{\mathrm{NP}}^i$ for each function, but in our analysis we finally opted  fo a single common value.\label{foot:1}}.

\begin{table}[h!]
    \centering
    \begin{tabular}{||c|c||}
    \hline
            & $\{\mathcal{C}_i,S_{\gamma i},\mathcal{M}_i\}$ eq.~(\ref{eq:Zexaxt}),~(\ref{eq:finp}) \\
         \hline
     $\beta_{\mathrm{NP}}^i$ (GeV$^{-1}$) & 2.5 \\
    \hline
    \end{tabular}
     \begin{tabular}{||c|c||}
    \hline
           & $\{\mathcal{C}_i, S_{\gamma i}, \mathcal{M}_i\}$  eq.~(\ref{eq:b*}) \\
         \hline
    $b_{\mathrm{max}}$ (GeV$^{-1}$)   & 2.5  \\
    \hline
    \end{tabular}
    \caption{\label{tab1}Values of non-perturbative parameter $\beta_{\mathrm{NP}}^i$ and $b_{\mathrm{max}}$ prescription chosen for collinear-soft, dijet soft and ${\cal M}$-functions (see also footnote~\ref{foot:1}). Indices $i=q,\bar q,g$. 
    }
\end{table}
A first solution for $\zeta$-prescription was  also provided in the same work.
The values of $\zeta_{1,0}$ and $\zeta_{2,0}$ have been calculated perturbatively and in exact form in \cite{Scimemi:2018xaf,Vladimirov:2019bfa,Scimemi:2019cmh} and \cite{delCastillo:2020omr} respectively.
%In the code ART23~\cite{Moos:2023yfa} and ART25~\cite{Moos:2025sal} the exact expression of  $\zeta_{1,0}$. 
The expression for $\zeta_{2,\mu_f}$ is
 \begin{align}
 \label{eq:Zexaxt}
     \zeta_{2,\mu_f}^{S_{\gamma q,\gamma g}}(b_T)=
     \zeta_{2,\mu_f}^{S_{\gamma q,\gamma g}\text{pert}} e^{-b_T^2/B_\zeta
     ^2}+
     \zeta_{2,\mu_f}^{S_{\gamma q,\gamma g}\text{exact}} (1-e^{-b_T^2/B_{\zeta}
     ^2})
 \end{align}
 with $B_{\zeta}=0.01$ GeV$^{-1}$ in ART23/ART25 and $B_{\zeta}=1$ GeV$^{-1}$ in \cite{delCastillo:2021znl}. 
 In this work we use $B_{\zeta}=0.01$ GeV$^{-1}$ for consistency with more modern versions of the code ART23 and ART25.
 Because in the cross section
 we have the product of 2 collinear-soft and one dijet soft functions treated independently   we refer to this way to implement the $\zeta$-prescription and "CCS-scheme".

Despite the fact that in CCS scheme the logs are resummed in each function separately, in practice the angular  $c_{\bmat{b}}$ dependence of the collinear-soft and soft  anomalous dimensions creates problems. As noticed in~\cite{delCastillo:2021znl} only angular $c_{\bmat{b}}$-independent parts of the anomalous dimensions can be resummed while for the rest one relies on a perturbative treatment. Moreover  the interval  of possible $\mu$-scale values  is limited in order not to spoil mathematical consistency.

An alternative approach that we follow here uses the ${\cal M}$-function introduced in sec.~\ref{sec:kernels}. This consists in calculating  the special initial saddle point scales of $\zeta$-prescription  for the ${\cal M}$-function.
 They are define by  $(\mu_0, \zeta_{2,0}^{\cal M})$  such that,
\begin{align}
    \mathcal{D}_i(\mu_0(b_T),b_T)=0, \quad  \gamma_{\mathcal{M}_i}(\mu_0(b_T),\zeta_{2,0}^{\mathcal{M}_i}(b_T))=0\,,\qquad \text{ with } i=q,\bar q, g\,.
\end{align}
To obtain the value of $\zeta_{2,0}^\mathcal{M}$ perturbatevely at one loop we start from the definition of the anomalous dimension of the total soft function for the gluon and quark channels in eq.~(\ref{eq:GMs}),
\begin{align}
  \gamma_{\mathcal{M}_g}^{[1]}
  &=- 4 C_A\ln\frac{\zeta_2}{ \zeta_{2,0}^{{\cal M}_g} }\,,\nn
  \\
    \gamma_{\mathcal{M}_q}^{[1]} 
    &= - 4C_F \ln\frac{\zeta_2}{  \zeta_{2,0}^{{\cal M}_q}}\,,
    \label{eq:gammaM}
\end{align}
where we have put 
\begin{align}
&\zeta_{2,0}^{{\cal M}_q} = \lp \frac{p_T^2 R^2 }{\hat{s}} \rp^{\frac{C_F + C_A}{C_F}} \lp \frac{\hat{t}}{\hat{u}} \rp^{\frac{C_F - C_A}{C_F}} \,,\nonumber \\
    &\zeta_{2,0}^{{\cal M}_g} = \lp \frac{p_T^2 R^2}{\hat{s}} \rp^{2\frac{C_F}{C_A}} \,.  
\end{align}
The evolution kernel of the total function $\cal M$ using the $\zeta$ prescription is then given by:

\begin{align}
{\cal R}_{{\cal M}_{ q, g}}:&\;  (\mu_0,\zeta_{2,0}^{{\cal M}_{q,g}}(b_T))=\left( \frac{2 e^{-\gamma_E}}{b^*},\zeta_{2,0}^{{\cal M}_{q,g}}(b_T)\right)\longrightarrow (p_T,\;1) \,,
\end{align}
and the problem of angular dependence   in the anomalous dimension disappears. 

In the ${\cal M}$ function the angular dependence is by definition only present   at the  initial scale where this function is perturbatively evaluated and the results at one loop are shown in the next-section.

In the proposed factorization it is necessary to introduce non-perturbative effects. In CCS-scheme we need to specify a non-perturbative part for  collinear-soft and dijet soft functions
\begin{align}
    \mathcal{C}_{i}(b_T, R; p_T) & = \mathcal{R}_{\mathcal{C}_{i}}(b^*,\mu_{\mathcal{C}_{i}}\rightarrow p_T)\mathcal{C}_i^{\it pert}
    (b^*, R; \mu_{\mathcal{C}_{i}}) f^{\mathrm{NP}}_{\mathcal{C}_{i}}(b_T), \nn\\
     S_{\gamma i}(b_T; p_T,1) & = \mathcal{R}_{S_i}(b^*,\{\mu_0,\zeta_{2,0}^{S_{\gamma i}}\}\rightarrow\{p_T,1\})S_{\gamma i}^{\it pert}(b^*; \mu_0, \zeta_{2,0}^{S_{\gamma i}})
      f^{\mathrm{NP}}_{\mathcal{S}_i}(b_T), \qquad \text{with }i=q, g
\end{align}
where the functions with label {\it pert} are their perturbative part in the $\overline{\rm MS}$ scheme, currently known at one loop.

The  $\mathcal{M}$-scheme needs a different set of non-perturbative parameters, namely
\begin{align}
\mathcal{M}_{i}(b_T,R; p_T, 1) & = \mathcal{R}_{\mathcal{M}_{i}}(b^*,\{\mu_0,\zeta_{2,0}^{\mathcal{M}_i}\}\rightarrow\{p_T,1\})\mathcal{M}^{\it pert}_{i}
    (b^*,R; \mu_0,\zeta_{2,0}^{\mathcal{M}_{i}}) f^{\mathrm{NP}}_{\mathcal{M}_{i}}(b_T),\qquad i=q, g\,.
    \end{align}
    %The calculation  of saddle point used in this  function is reported in appendix \ref{app:Msadp}. \textcolor{red}{Patri: but in A.1 we do NOT calculate the saddle point, but the solution of $\zeta_2$ derived from it, with its perturbative and exact solution. We use this expression below eq. (6.9).}
The non-perturbative model is given by
\begin{equation}
\label{eq:finp}
    f^{\mathrm{NP}}_{i}(b_T)=\exp \lp -\frac{b_T^2}{(\beta_{\mathrm{NP}}^i)^2} \rp,\quad\quad i=\mathcal{C}_{q, g},\mathcal{M}_{q, g}, S_{\gamma q, \gamma g}.
\end{equation}
The values of $\beta_{\mathrm{NP}}^i$ determine the non-perturbative model.
Assuming that  most of the contribution is of perturbative origin   the values of 
$\beta_{\mathrm{NP}}^i$ should be around 1-3 GeV$^{-1}$. We have finally made the choice of tab.~\ref{tab1} to provide numerical results.

\subsection{Final expression of the unpolarized hadronic tensor }
Using the ${\cal M}$-scheme the hadronic tensor contributions are respectively
\begin{align}
W_{g}^{UU}     &=  r_T \sigma_0^{gU}\,\sum_{q=\text{flav.}} H^{U}_{\gamma^*g\to  q \bar q}(\hat{s},\hat{t}, \hat{u},\mu)   \int \frac{b_T db_T}{(2\pi)}  J_0(r_T b_T) J_{\bar q}(p_T,R,\mu_{J_{\bar q}})      
    J_{q}(p_T,R,\mu_{J_q})   f_{1}^{ g}(\xi, b_T)\nn \\
    &  \times 
    {\cal R}_{F}^g(\{\mu_{0},\zeta_{1,0}^g\} \rightarrow \{p_T,p_T^2\})
    {\cal R}_{\mathcal{M}_g}(\{\mu_0,\zeta_{2,0}^{{\cal M}_g} \} \rightarrow \{p_T,1 \})
    {\cal R}_{J_{\bar q}}(\mu_{J_{\bar q}} \rightarrow p_T)
    {\cal R}_{J_{q}}(\mu_{J_q} \rightarrow p_T) \nn \\
    & \times\int_{-\pi}^{\pi} d\phi_b
     \mathcal{M}_{ g}(\bmat{b}_\perp,R,\mu_0,\zeta_{2,0}^{{\cal M}_g}) ,
     \\
    W_{q}^{UU} 
    &=  r_T\sigma_0^{qU}\, H^{U}_{\gamma^*q\to g q }(\hat{s},\hat{t}, \hat{u},\mu)   \int \frac{b_T db_T}{(2\pi)}  J_0(r_T b_T) J_{g}(p_T,R,\mu_{J_g})      
    J_{q}(p_T,R,\mu_{J_q})   f_{1}^q\xi, b_T)\nn \\
    &  \times 
    {\cal R}_{F}^q(\{\mu_{0},\zeta_{1,0}^q\} \rightarrow \{p_T,p_T^2\})
    {\cal R}_{\mathcal{M}_q}(\{\mu_0,\zeta_{2,0}^{\mathcal{M}_q} \} \rightarrow \{p_T,1 \})
    {\cal R}_{J_{g}}(\mu_{J_g} \rightarrow p_T)
    {\cal R}_{J_{q}}(\mu_{J_q} \rightarrow p_T) \nn \\
    & \times\int_{-\pi}^{\pi} d\phi_b
     \mathcal{M}_{ q}(\bmat{b}_\perp,R,\mu_0,\zeta_{2,0}^{{\cal M}_q}) ,  
     \end{align}
 where  $J_0$ is the Bessel function and similarly one can define $W_{\bar q}^{UU}$ interchanging $q\leftrightarrow \bar q$.
If one follows the scale fixing according the work~\cite{delCastillo:2021znl} 
 the evolution factors of the collinear-soft and dijet soft functions have angular dependence and the formulas of the hadronic tensor change.  Using  eq.~(\ref{eq:r1}-\ref{eq:r2}) one writes
\begin{align}
W_{g}^{UU} 
    &= r_T \sigma_0^{gU}\,\sum_{q=\text{flav.}} H^{U}_{\gamma^*g\to  q \bar q}(\hat{s},\hat{t}, \hat{u},\mu)   \int \frac{b_T db_T}{(2\pi)}  J_0(r_T b_T) J_{\bar q}(p_T,R,\mu_{J_{\bar q}})      
    J_{q}(p_T,R,\mu_{J_q})   f_{1}^{ g}(\xi, b_T)\nn \\
    &  \times \int_{-\pi}^{\pi} d\phi_b 
    {\cal R}_g\lp (\{\mu_k\},\zeta_{1,0}^g,\zeta_{2,0}^{S_{\gamma g}}) \to (p_T,p_T^2,1) \rp 
    \nn \\
    & \times
     \mathcal{C}_{ \bar q}(\bmat{b}_\perp,R,\mu_{{\cal C}_{\bar q}}) 
     \mathcal{C}_{ q}(\bmat{b}_\perp,R,\mu_{{\cal C}_q})
       \mathcal{S}_{ \gamma g}(\bmat{b}_\perp,\mu_0,\zeta_{2,0}^{S_{\gamma g}})
     \\
    W_{q}^{UU}
    &= r_T \sigma_0^{qU}\, H^{U}_{\gamma^*q\to g q }(\hat{s},\hat{t}, \hat{u},\mu)   \int \frac{b_T db_T}{(2\pi)}  J_0(r_T b_T) J_{g}(p_T,R,\mu_{J_g})      
    J_{q}(p_T,R,\mu_{J_q})   f_{1}^q\xi, b_T)\nn \\
    &  \times \int_{-\pi}^{\pi} d\phi_b
    {\cal R}_q\lp (\{\mu_k\},\zeta_{1,0}^q,\zeta_{2,0}^{S_{\gamma q}}) \to (p_T,p_T^2,1) \rp
    \nn \\
    & \times
      \mathcal{C}_{ g}(\bmat{b}_\perp,R,\mu_{{\cal C}_g}) 
     \mathcal{C}_{q}(\bmat{b}_\perp,R,\mu_{{\cal C}_q})
       \mathcal{S}_{ \gamma q}(\bmat{b}_\perp,\mu_0,\zeta_{2,0}^{S_{\gamma q}}),  
     \end{align}
and  the angular integration includes also the evolution factors.

\subsection{Final expression of the hadronic tensor for Sivers function measurement}
The general expression for the hadronic tensor has been given in formula~(\ref{eq:wut}) without specifying the evolution  of the distributions that compose it. 
For the case in which we use the ${\cal M}$-scheme we have
\begin{align}
W_{g}^{UT} 
    &=- r_T S_T\sigma_0^{gU}\,\sum_{q=\text{flav.}} H^{U}_{\gamma^*g\to  q \bar q}(\hat{s},\hat{t}, \hat{u},\mu)   \int \frac{b_T db_T}{(2\pi)}  J_1(r_T b_T) J_{\bar q}(p_T,R,\mu_{J_{\bar q}})      
    J_{q}(p_T,R,\mu_{J_q})   b_T Mf_{1T}^{\perp g}(\xi, b)\nn \\
    &  
     \times 
    {\cal R}_{F}^g(\{\mu_{0},\zeta_{1,0}^g\} \rightarrow \{p_T,p_T^2\})
    {\cal R}_{\mathcal{M}_g}(\{\mu_0,\zeta_{2,0}^{{\cal M}_g} \} \rightarrow \{p_T,1\})
    {\cal R}_{J_{\bar q}}(\mu_{J_{\bar q}} \rightarrow p_T)
    {\cal R}_{J_{q}}(\mu_{J_q} \rightarrow p_T) \nn \\
    & \times\int_{-\pi}^{\pi} d\phi_b\sin^2(\phi_S - \phi_b) 
     \mathcal{M}_{ g}(\bmat{b}_\perp,R,\mu_0,\zeta_{2,0}^{{\cal M}_g}) ,
     \\
    W_{q}^{UT} 
    &=- r_T S_T\sigma_0^{qU}\, H^{U}_{\gamma^*q\to g q }(\hat{s},\hat{t}, \hat{u},\mu)   \int \frac{b_T db_T}{(2\pi)}  J_1(r_T b_T) J_{g}(p_T,R,\mu_{J_g})      
    J_{q}(p_T,R,\mu_{J_q})   b_T Mf_{1T}^{\perp q}(\xi, b_T)\nn \\
    &  
   \times 
    {\cal R}_{F}^q(\{\mu_{0},\zeta_{1,0}^q\} \rightarrow \{p_T,p_T^2\})
    {\cal R}_{\mathcal{M}_q}(\{\mu_0,\zeta_{2,0}^{{\cal M}_q} \} \rightarrow \{p_T,1 \})
    {\cal R}_{J_{g}}(\mu_{J_g} \rightarrow p_T)
    {\cal R}_{J_{q}}(\mu_{J_q} \rightarrow p_T) \nn \\
    & \times\int_{-\pi}^{\pi} d\phi_b\sin^2(\phi_S - \phi_b) 
     \mathcal{M}_{ q}(\bmat{b}_\perp,R,\mu_0,\zeta_{2,0}^{{\cal M}_q}) ,  
     \end{align}
 and the expression holds for $W^{UT}_{\bar q}$ with the interchange $q\leftrightarrow \bar q$.
     For completeness we provide the one-loop expression for ${\cal M}$ function in the next section.

If one follows the scale fixing according the work~\cite{delCastillo:2021znl} 
 the evolution factors of the CCS-scheme have angular dependence and the formulas of the hadronic tensor change to
     \begin{align}
W_{g}^{UT} 
    &=- r_T S_T\sigma_0^{gU}\,\sum_{q=\text{flav.}} H^{U}_{\gamma^*g\to  q \bar q}(\hat{s},\hat{t}, \hat{u},\mu)   \int \frac{b_T db_T}{(2\pi)}  J_1(r_T b_T) J_{\bar q}(p_T,R,\mu_{J_{\bar q}})      
    J_{q}(p_T,R,\mu_{J_q})  b_T M f_{1T}^{\perp g} (\xi, b_T)\nn \\
    &  \times \int_{-\pi}^{\pi} d\phi_b \sin^2(\phi_S - \phi_b) 
    {\cal R}_g\lp (\{\mu_k\},\zeta_{1,0}^g,\zeta_{2,0}^{S_{\gamma q}}) \to (p_T,p_T^2,1) \rp 
    \nn \\
    & \times
     \mathcal{C}_{ \bar q}(\bmat{b}_\perp,R,\mu_{{\cal C}_{\bar q}}) 
     \mathcal{C}_{ q}(\bmat{b}_\perp,R,\mu_{{\cal C}_q})
       \mathcal{S}_{ \gamma g}(\bmat{b}_\perp,\mu_0,\zeta_{2,0}^{S_{\gamma g}})
     \\
    W_{q}^{UT}
    &= - r_T S_T\sigma_0^{qU}\,\sum_{q=\text{flav.}} H^{U}_{\gamma^*q\to g q }(\hat{s},\hat{t}, \hat{u},\mu)   \int \frac{b_T db_T}{(2\pi)}  J_0(r_T b_T) J_{g}(p_T,R,\mu_{J_g})      
    J_{q}(p_T,R,\mu_{J_q})   b_T M f_{1T}^{\perp q}(\xi, b_T)\nn \\
    &  \times \int_{-\pi}^{\pi} d\phi_b\sin^2(\phi_S - \phi_b) 
    {\cal R}_q\lp (\{\mu_k\},\zeta_{1,0}^q,\zeta_{2,0}^{S_{\gamma q}}) \to (p_T,p_T^2,1) \rp
    \nn \\
    & \times
      \mathcal{C}_{ g}(\bmat{b}_\perp,R,\mu_{{\cal C}_g}) 
     \mathcal{C}_{q}(\bmat{b}_\perp,R,\mu_{{\cal C}_q})
       \mathcal{S}_{ \gamma q}(\bmat{b}_\perp,\mu_0,\zeta_{2,0}^{S_{\gamma q}}),  
     \end{align}
the angular integration includes also the evolution factors and we have used eq.~(\ref{eq:r1}-\ref{eq:r2}).
As in the unpolarized case  in the latter formulas
     the resummation of  different functions   is fixed  with a different choice of scales, which minimizes the logs of each function separately.

\subsection{$\cal M$-function scheme at one loop}
In this section we collect the relevant one-loop expressions for the ${\cal M}$-function.
Using the notation of eq.~(\ref{eq:gMexpanded})
At one loop  we have
\begin{align}
    \mathcal{M}_g(\bmat{b}_\perp,R,\mu,\zeta_2)^{[1]} &= \mathcal{C}_{\bar{q}}(\bmat{b}_\perp, R,\mu)^{[1]} + \mathcal{C}_{q}(\bmat{b}_\perp, R,\mu)^{[1]} + S_{\gamma g}(\bmat{b}_\perp,\mu,\zeta_2)^{[1]}
\\
    \mathcal{M}_q(\bmat{b}_\perp,R,\mu,\zeta_2)^{[1]} &= \mathcal{C}_{g}(\bmat{b}_\perp, R,\mu)^{[1]} + \mathcal{C}_{q}(\bmat{b}_\perp, R,\mu)^{[1]} + S_{\gamma,f}(\bmat{b}_\perp,\mu,\zeta_2)^{[1]}
\end{align}
where on the r.h.s. we indicate the one-loop expression for all these functions. Combining the results of previous calculations and being $B=|\bmat{b}_\perp|^2/4=b_T^2/4$ we have 
\begin{align}
\mathcal{M}_{ g}^{[0]}(\bmat{b}_\perp,R,\mu,\zeta_2) &=1\,,\\ \nn
\mathcal{M}_{ g}^{[1]}(\bmat{b}_\perp,R,\mu,\zeta_2) &=   a_s\lbc 
 C_F \lb \frac{\pi^2}{3}  + 2 \ln^2 \lp \frac{ B e^{2\gamma_E} \mu^2}{ -A_{\bmat{b}}}\rp + 4 \text{Li}_2(1+A_{\bmat{b}})  \rb 
 \\[5pt] & \nn
 +C_A \lb  - 2 \ln (B e^{2\gamma_E} \mu^2)   \ln \zeta_2  - \ln^2(-A_{\bmat{b}}) - \frac{\pi^2}{3} - 2  \text{Li}_2(1+A_{\bmat{b}}) \rb 
\rbc \\[5pt] &
 - 8 a_sC_F \Bigg[\ln^2\left(\frac{\mu e^{\gamma_E} b_T}{R}\right) + \frac{\pi^2}{8} + \ln^2(-i\cos\phi_b) + 2\ln\left(\frac{\mu e^{\gamma_E} b_T}{R}\right)\ln(-i\cos\phi_b) \Bigg] +{\cal O}(a_s^2),
\end{align}
for gluons and
\begin{align}
 \mathcal{M}_q^{[0]} (\bmat{b}_\perp,R,\mu,\zeta_2) &=1\,,\\
\nn
 \mathcal{M}_q^{[1]} (\bmat{b}_\perp,R,\mu,\zeta_2) &=  a_s \lbc
C_A \lb  \frac{\pi^2}{6}  +  \ln^2 \lp \frac{B\mu^2 e^{2\gamma_E}}{ -A_{\bmat{b}_\perp}}\rp + 2 \text{Li}_2(1+A_{\bmat{b}_\perp}) + 2\ln( B\mu^2 e^{2\gamma_E})   \ln \frac{(n \cdot v_1) (\bmat{v}_2 \cdot \bmat{b}_\perp)}{(n \cdot v_2) (\bmat{v}_1 \cdot \bmat{b}_\perp)}   \rb 
\\[5pt] & \hspace{-2cm}
+ C_F \ln( B\mu^2 e^{2\gamma_E}) \lb \ln (B\mu^2 e^{2\gamma_E}) -2 \ln \zeta_2 + 2 \ln \lp\frac{2  (n \cdot v_2)}{(v_1\cdot v_2) (n \cdot v_2)} \rp  -\frac{\pi^2}{6} + 4\,\text{ln} (-i \,\bmat{v}_1 \cdot \hat{\bmat{b}}_\perp) \rb \nn
 \rbc
 \\[5pt] &  \hspace{-2cm}
 - 4a_s(C_A +C_F) \Bigg[\ln^2\left(\frac{\mu e^{\gamma_E} b_T}{R}\right) + \frac{\pi^2}{8} + \ln^2(-i\cos\phi_b) + 2\ln\left(\frac{\mu e^{\gamma_E} b_T}{R}\right)\ln(-i\cos\phi_b) \Bigg]\,,
\end{align}
for quarks, where we have used $\hat{\bmat{b}}_\perp = \bmat{b}_\perp/|\bmat{b}_\perp|$. 
In the hadronic tensor these expressions are angularly integrated providing
\begin{align}
 \label{eq:M1g} \nn
    \int_{-\pi}^{\pi} d\phi_b \sin^2(\phi_S-\phi_b) \mathcal{M}_{g}^{[1]}(\textbf{b}_\perp,R,\mu,\zeta_2) & \\[5pt]\nn
    &\hspace{-5cm}
    =  a_s\lbc 
    C_F \lb \frac{\pi^3}{3}  
    + 2\pi \ln^2 \lp B \mu^2 e^{2\gamma_E}\rp    + 
    2 I_{A_b^2}
    - 4 \ln \lp B \mu^2 e^{2\gamma_E}\rp I_{A_b}
    + 4 I_{Li_2}\rb 
    \\[5pt] &\hspace{-5cm}\nn
    +C_A \lb  - 2\pi \ln (B \mu^2 e^{2\gamma_E})   \ln \zeta_2   - I_{A_b^2} - \frac{\pi^3}{3} - 2 I_{Li_2} \rb 
    \rbc  \\[5pt]  \hspace{-3cm} &\hspace{-5cm}
     -8 a_s C_F\Bigg[\pi\ln^2\left(\frac{\mu e^{\gamma_E}b_T}{R}\right)  + \frac{\pi^3}{8} + 2\ln \left(\frac{\mu e^{\gamma_E}b_T}{R}\right) I_{log} + I_{log2} \Bigg]\,,
\end{align}
for gluons and
\begin{align} 
\label{eq:M1q}\nn
    \int_{-\pi}^{\pi} d\phi_b \sin^2(\phi_S-\phi_b)  \mathcal{M}_q^{[1]} (\bmat{b}_\perp,R,\mu,\zeta_2) &= a_s \lbc
C_A \lb  \frac{\pi^3}{6}  
+ \pi \ln^2 \lp B\mu^2 e^{2\gamma_E}\rp 
+ I_{A_b^2} 
\\[5pt]&\hspace{-5cm}\nn
 -2 \ln \lp B\mu^2 e^{2\gamma_E}\rp I_{A_b}
+ 2 I_{Li_2} + 2\ln( B\mu^2 e^{2\gamma_E}) \ln\frac{t}{u} \lp  -\cos 2\phi_S I_0(1) + \cos^2\phi_S I_0(0) \rp  \rb 
\\[5pt]&\hspace{-5cm}\nn
+ C_F \ln( B\mu^2 e^{2\gamma_E}) \lb\pi \ln (B\mu^2 e^{2\gamma_E})  -2 \pi\ln \zeta_2   -\frac{\pi^3}{6} 
\\[5pt] &\hspace{-5cm}
-2 \cos2\phi_S \ln\frac{t}{u} I_0(1)  + 2\cos^2\phi_S \ln\frac{t}{u} I_0(0) - 2I_{A_b} \rb 
 \rbc\nn
 \\[5pt]&\hspace{-5cm}
-4 a_s (C_A + C_F) \Bigg[\pi\ln^2\left(\frac{\mu e^{\gamma_E} b_T}{R }\right)  + \frac{\pi^3}{8} + 2\ln \left(\frac{\mu e^{\gamma_E} b_T}{R }\right) I_{log} + I_{log2} \Bigg]\,,
\end{align}
for quarks.
The explicit
expression for the $I$ terms in these equations is provided in appendix \ref{app:angularintegrals}.

%%%%%%%%%%%%%%%%%%
\section{The quark and gluon CS-kernels for evolution}
\label{sec:CSkernel}
%%%%%%%%%%%%%%%%%%%%
In the ART23/ART25 extractions of unpolarized TMD the  the evolution (CS) kernel has the general form
\begin{eqnarray}\label{def:CS-kernel}
\mathcal{D}_{q,g}(b_T,\mu)=\mathcal{D}_{{q,g},\text{pert}}(b^*,\mu^*)+\int_{\mu^*_{q,g}}^\mu \frac{d\mu'}{\mu'}\Gamma_{\text{cusp}}(\mu')+\mathcal{D}_{q,g,\text{NP}}(b_T),
\end{eqnarray}
where $b^*$ ,
\begin{eqnarray}\label{CS:scale}
b^*_{q,g}(b_T)=\frac{b_T}{\sqrt{1+\frac{ b_T^2}{(B^{q,g}_{\text{NP}})^2}}},\qquad \mu^*_{q,g}(b_T)=2\exp^{-\gamma_E}/b^*_{q,g}(b_T)\,,
\end{eqnarray}
and $B_{\text{NP}}^{q,g}$'s for the kernels take a value as in tab.~\ref{tab:CS},

 $\mathcal{D}_{\text{pert}}$'s are the perturbative expression of the kernel at N$^3$LO \cite{Moult:2022xzt, Duhr:2022yyp}, and $\mathcal{D}_{\text{NP}}$'s are the non-perturbative parts to be modeled. 
 Up to three loops we have 
  ${\cal D}_{pert}^q=(C_F/ C_A){\cal D}_{pert}^g$
 and $\Gamma^q_{\text{cusp}}=(C_F/ C_A) \Gamma^g_{\text{cusp}}$ and differences appear at 4 loops.
 The integral term in eq.~(\ref{def:CS-kernel}) describes the evolution from the scale $\mu^*_q$ to $\mu$. 
 For the quark case we have
\begin{eqnarray}\label{CS:NP-part}
\mathcal{D}_{\text{NP}}^q(b_T)=b_Tb^*\left[c_0+c_1\ln \left(\frac{b^*}{B_{\text{NP}}^q}\right)\right],
\end{eqnarray}
with $c_0$ and $c_1$ free parameters.  At large-$b_T$ the $b^*$ prescription freezes  at 
 $b_T \sim B_{\text{NP}}^q$ preventing the approaching to the Landau pole of the perturbative part.
 and the kernel is dominated by $\mathcal{D}_{\text{NP}}$. 
 The two fits use different sets of data and the phenomenologically extracted parameters have some tension between each other, as one can see in tab.~\ref{tab:CS}.
 \begin{table}[h]
     \centering
     \begin{tabular}{|c|c|c|}
     \hline\hline
     Parameter     & ART25 & ART23 \\
     \hline
     $c_0$  (GeV$^{-2}$)   & $0.0859^{+0.0023}_{-0.0017}$ & $0.0369^{+0.0065}_{-0.0061}$
     \\
     $c_1$ (GeV$^{-2}$) & $0.0303^{0.0038}_{-0.0041}$ &
     $0.0582^{+0.0064}_{-0.0088}$\\
     $B_{NP}^q=B_{NP}^g$ (GeV$^{-1}$) &  1.5 (fixed,\cite{Moos:2025sal}) &  $1.56^{+0.013}_{-0.09}$ \cite{Moos:2023yfa}
     \\
     \hline\hline
  %   \multicolumn{2}{}{}
  %   For CCS scheme we use $B_{NP}^q=B_{NP}^g =1 \text{ GeV}^{-1}$ 
     \end{tabular}
     \caption{Parameters for the non-perturbative part of the Collins-Soper kernel, $\mathcal{D}_{\text{NP}}$ in ART25/ART23.  See discussion in the sec.~\ref{sec:CSkernel} .
     \label{tab:CS}}
 \end{table}
 
 The evolution kernel values are however different for quark and gluon cases so that we have to distinguish
 $\mathcal{D}^q$ and $\mathcal{D}^g$.
 In ART23/ART25 it was implemented also ${\cal D}_{NP}^q={\cal D}_{NP}^g$
  while here we prefer to set ${\cal D}_{NP}^q=(C_F/ C_A){\cal D}_{NP}^g$.
 The fits however are not very sensitive to the gluon non-perturbative parts (in fact they use DY and SIDIS data which are dominated by quark channels).

We find a great sensitivity to both quark and gluon evolution kernels. 
In $\mathcal{M}$-scheme we use the same values of the non-perturbative parameters as in the original codes (see tab. \ref{tab:CS}).
In CCS-scheme instead, we have less convergence and we choose the evolution kernel parameters $B_\text{NP}^q=B_\text{NP}^g=1$ GeV$^{-1}$, while keeping the rest of parameters ($c_0$ and $c_1$) as in tab. \ref{tab:CS}. 
I addition, we also need to improve the numerical stability by freezing the value of $\mathcal{D}_\text{NP}$ at $b=6$ GeV$^{-1}$.

%%%%%%%%%%%%%%%%%%%%%%%
\section{ Sivers function models}
\label{sec:SiversModels}
%%%%%%%%%%%%%%%%%%%%%%%%
In the past, several models for Sivers functions have been proposed for the quark distributions~\cite{Bury:2021sue,Bury:2020vhj,Echevarria:2020hpy}, while there are less studies for the gluon functions~\cite{DAlesio:2015fwo,Bacchetta:2020vty}. A comparison among models is difficult because of the methodology of evolution used in each work which is 
recapitulated
in tab.~\ref{tab:Siversmodels}.
The main differences come from the inclusion of the evolution and TMD fragmentation functions in the analysis of SIDIS data.In the references listed in tab.~\ref{tab:siv} there is a big difference in the functions extracted including or not including evolution factors. For consistency we only consider the extraction where evolution is included at the same order that we work.

\begin{table}[h]
    \centering
    \begin{tabular}{||c|c||}
    \hline
     Reference    & TMD Evolution kernel\\
     \hline
   \cite{DAlesio:2015fwo}      & none\\
   \cite{Echevarria:2020hpy} & NNLL (Collins like)\\
   \cite{Bury:2021sue,Bury:2020vhj} & N3LO (\texttt{artemide})\\
   \hline
    \end{tabular}
    \caption{Some models for extracted Sivers function and main features of the fits\label{tab:siv}}
    \label{tab:Siversmodels}
\end{table}
%%%%%%
In the present work we adopted the model and the parameters for the  quark Sivers function of ref.~\cite{Bury:2021sue} as default.
This model has been tested  in ~\cite{Bury:2021sue} to substantially agree with the one in \cite{Echevarria:2020hpy}.
There is no direct extraction of the Sivers gluon function up to now so that some ansatz must be provided.
In the extraction~\cite{Bury:2020vhj,Bury:2021sue} the flavor dependence is partially included and light quarks are distinguished from the sea. We consider the gluon contribution as part of the sea for those cases, so that  for the Sivers gluon function we use as default the same model as for the sea case  in~\cite{Bury:2021sue}.
In the next section we also study different normalization of this model in order to provide a discussion about this ansatz.
Here we provide some information about the quark models in ref.~\cite{Bury:2021sue}.
A summary of the model in \cite{Echevarria:2020hpy} is reported in the appendix \ref{sec:ekt}. 

\subsection{Models for quark Sivers function  }
By comparison  with models of ref.~\cite{Bury:2021sue} (BPV)
 and ref.~\cite{Echevarria:2020hpy} (EKT)  we obtain
\begin{equation}
    f_{1T}^{\perp q}(\xi,b_T)= f_{1T,\text{BPV}}^{\perp}(\xi,b_T)= f_{1T,\text{EKT}}^{\perp}(\xi,b_T)\,.
\end{equation}
where in the l.h.s we have our definition.

\subsubsection{Model BPV}

In \cite{Bury:2021sue} the auhtors consider the Sivers function as a generic non-perturbative function that is extracted from the data. The resulting definition is:

\begin{eqnarray}\label{def:model}
f_{1T,\text{BPV}}^\perp(\xi,b_T)= N_q \frac{(1-\xi) \xi^{\beta_q} (1+\epsilon_q \xi)}{n(\beta_q,\epsilon_q)}\exp\left(-\frac{r_0+\xi r_1}{\sqrt{1+r_2 \xi^2 b^2_T}}b^2_T\right),
\end{eqnarray}
where $n(\beta,\epsilon)= (3+\beta+\epsilon+\epsilon \beta)\Gamma(\beta+1)/\Gamma(\beta+4)$, such that
\begin{eqnarray}
\int_0^1 dx f_{1T,\text{BPV}}^\perp(\xi,0)= N_q .
\end{eqnarray}
The factor in front of the exponential is independent of the transverse variable $b_T$ and parametrizes the Qiu-Sterman function to be obtained in the limit $b_T\rightarrow 0$.
The values obtained  for the parameters in the fit are reported in appendix \ref{app:bury} table \ref{tab:bury}.

%%%%%%%%%%%%%%%%%
\section{Results}
\label{sec:results}
%%%%%%%%%%%%%%%%

\begin{figure}[h]
    \centering
    \includegraphics[width=0.30\linewidth]{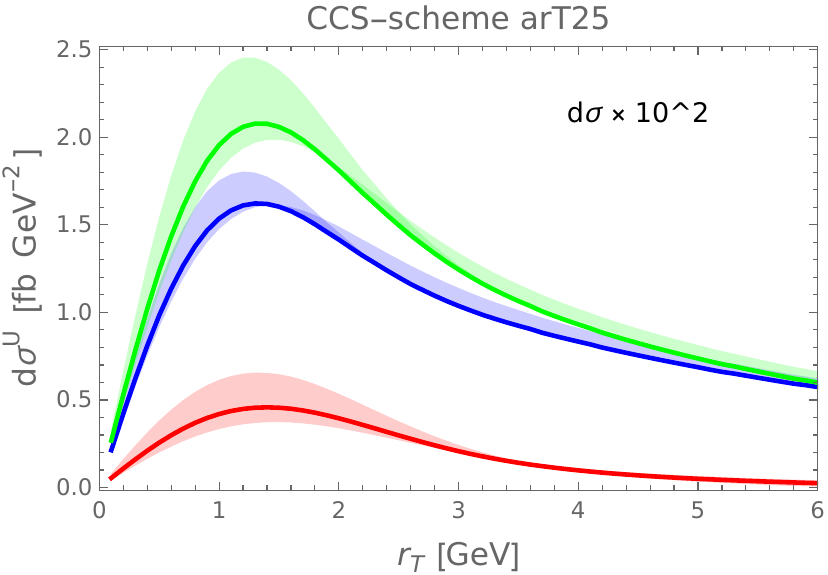}
\includegraphics[width=0.30\linewidth]{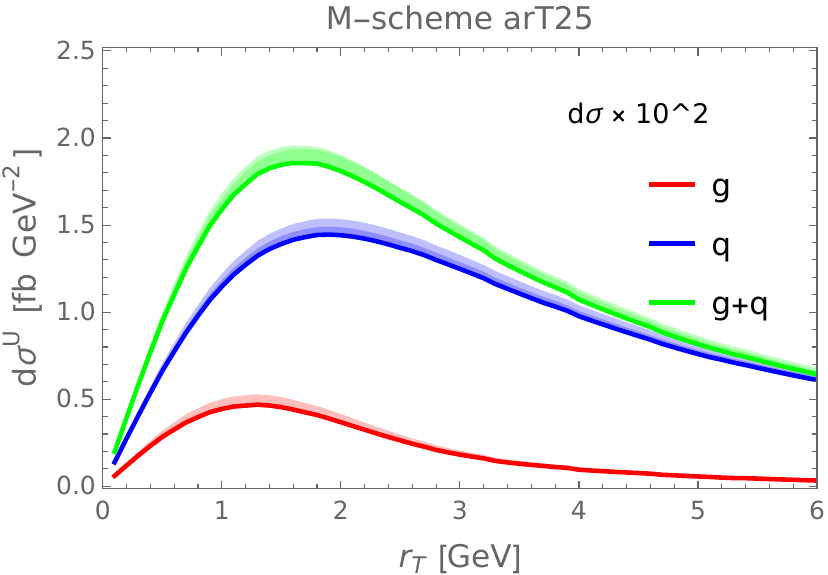}
\includegraphics[width=0.30\linewidth]{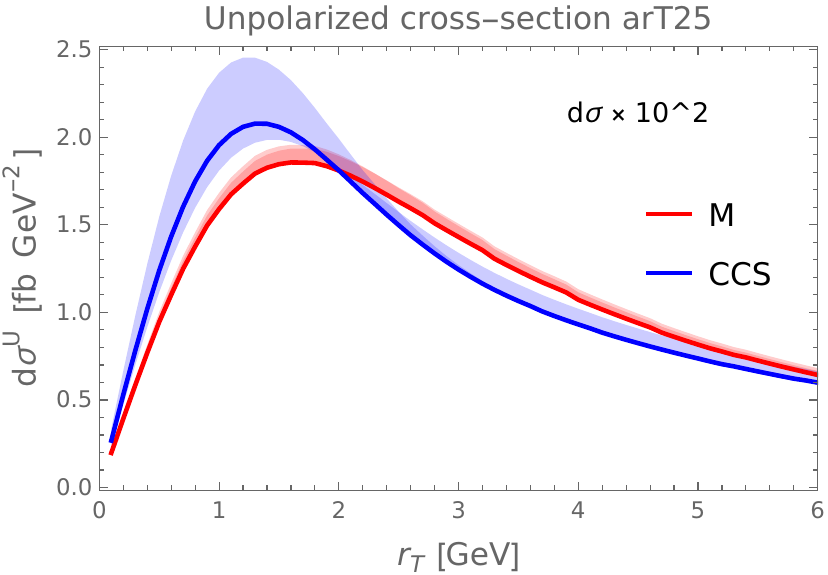}
    \caption{Unpolarized cross-section using CCS-scheme (left) and $\mathcal{M}$-scheme.
    (center). We show separately gluon-channel  (red), quark-channel contributions and final result (green). In the right plot we show the comparison of unpolarized cross-section in $\mathcal{M}$- and CCS-schemes, respectively, in red and blues lines. The band corresponds to hard function uncertainty. }
    \label{fig:Xsecs}
\end{figure}

The results presented in this work correspond to the cross section defined in eq.~(\ref{eq:tutxi}), evaluated at central rapidity $\eta_1=\eta_2=0$ using eq.~(\ref{eq:WUUUT}), $p_T=20$~GeV and a center of mass energy of $\sqrt{s}=140$~GeV. For central rapidity we have $x\in(0.0859, 0.5)$ and $\xi\in(0.1718,1)$ which corresponds to the inelasticity interval $y\in (0.16, 0.95)$. We use ART25~\citep{Moos:2025sal} settings as default.
Following the discussion in the previous sections, we present predictions in two renormalization schemes: one based on the $\mathcal{M}$ function (hereafter referred to as the $\mathcal{M}$ scheme), and another in which the $\mathcal{M}$ function is further factorized into collinear-soft and soft functions (the CCS scheme). The unpolarized dijet cross sections obtained in the two schemes are shown in Fig.~\ref{fig:Xsecs}. The CCS scheme was previously employed in~\citep{delCastillo:2021znl}. After revisiting the numerical implementation used in that analysis, we identified and corrected sign inconsistencies; however, the overall conclusions remain consistent with those reported there. 

The two schemes lead to compatible predictions for the central values of the distributions, and the relative contributions from quark and gluon channels are found to be consistent. The uncertainty bands shown in the figures are obtained from varying the hard scale in the interval $\mu\in (1/2,2)p_T$ and represent the dominant perturbative uncertainty, in agreement with~\citep{delCastillo:2021znl}.

\begin{figure}[h]
    \centering
    \includegraphics[width=0.30\linewidth]{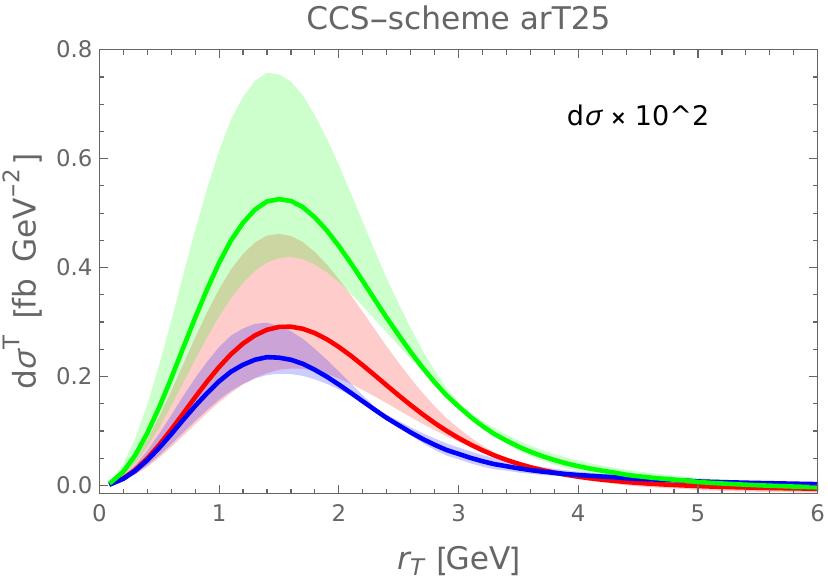}
\includegraphics[width=0.30\linewidth]{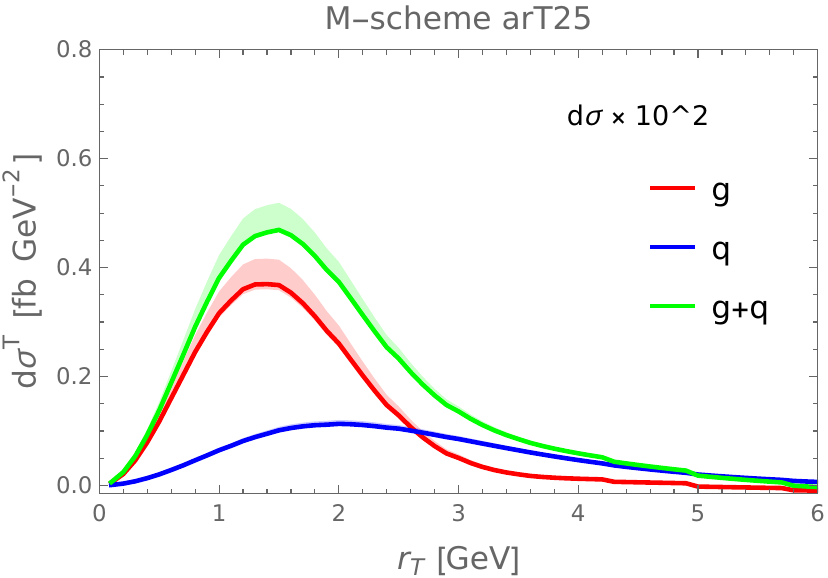}
\includegraphics[width=0.30\linewidth]{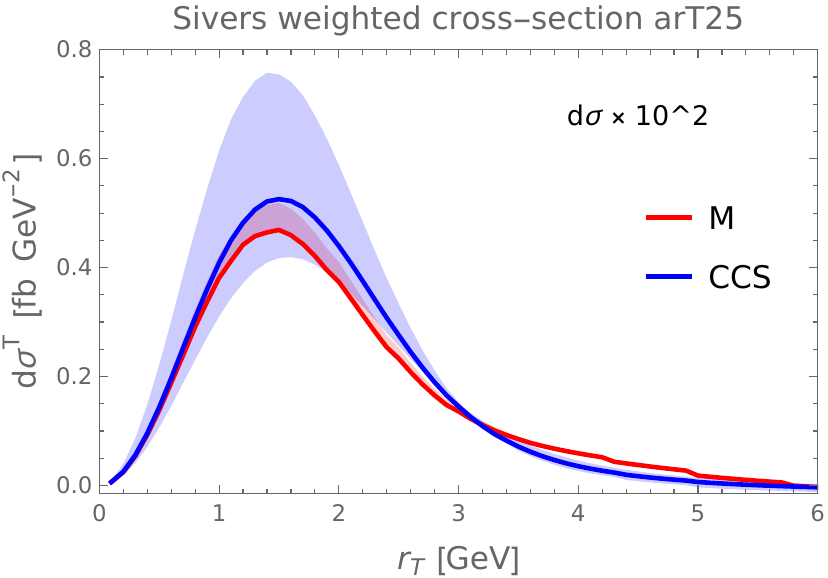}
    \caption{Sivers weighted cross-section using CCS-scheme (left) and $\mathcal{M}$-scheme.
    (center). We show separately gluon-channel  (red), quark-channel contributions and final result (green). In the right plot we show the comparison of Sivers weighted cross-section in $\mathcal{M}$- and CCS-schemes, respectively, in red and blues lines. The band corresponds to hard function uncertainty.}
    \label{fig:Xsecsp}
\end{figure}

The uncertainty is larger in the CCS scheme than in the $\mathcal{M}$ scheme, which we interpret as an indication of improved perturbative stability in the latter. The comparison of the final cross sections obtained in the two schemes is displayed on the right-hand side of Fig.~\ref{fig:Xsecs}.
%%%%%%%%%%%%%%%%%%%
%%%%%%%%%%%%%%%%%%%%%%%%
Similar observations are true for the polarized cross sections shown in fig.~\ref{fig:Xsecsp}. In this case quark and gluon distributions contribute almost equally to the cross section for the CCS-scheme, while their separate contribution is different in the case of the ${\cal M}$-scheme.  The two schemes however agree within errors and the final result of the polarized cross section  are compared on r.h.s of fig.~\ref{fig:Xsecsp} with bands corresponding to hard scale errors.

%%%%%%%%%%%%%%%%%%%%%%%%%%
\begin{figure}[h]
    \centering
    \includegraphics[width=0.5\linewidth]{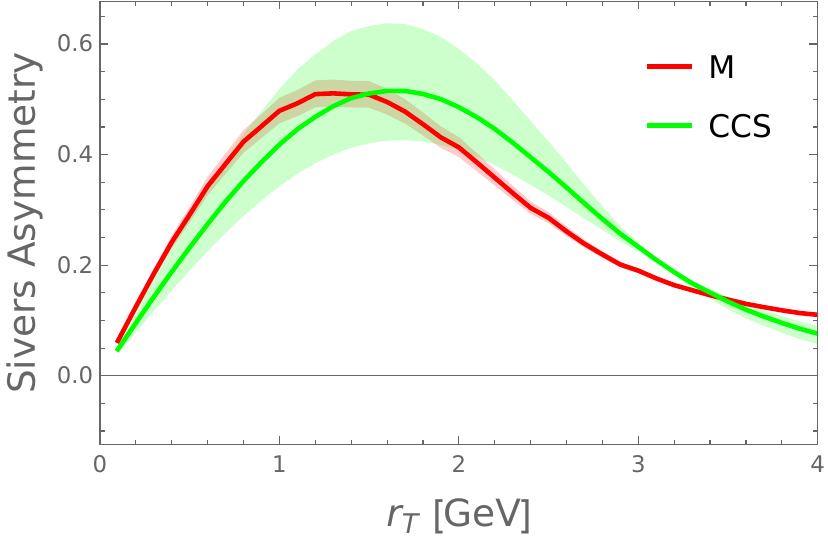}
    \caption{Sivers asymmetry in CCS-scheme (green) and $\mathcal{M}$-scheme (red). }
    \label{fig:Xsecsa}
\end{figure}

\begin{figure}[h]
    \centering
    \includegraphics[width=0.5\linewidth]{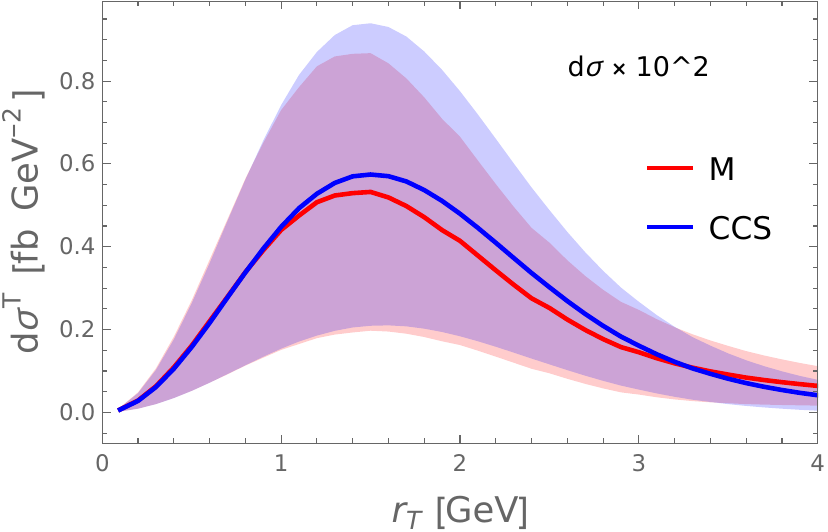}
    \caption{Sivers weighted cross-section with error of the Sivers model in $\mathcal{M}$- and CCS-scheme, respectively in red and blue lines. }
    \label{fig:Xsecsaerr}
\end{figure}

\begin{figure}[h]
    \centering
    \includegraphics[width=0.5\linewidth]{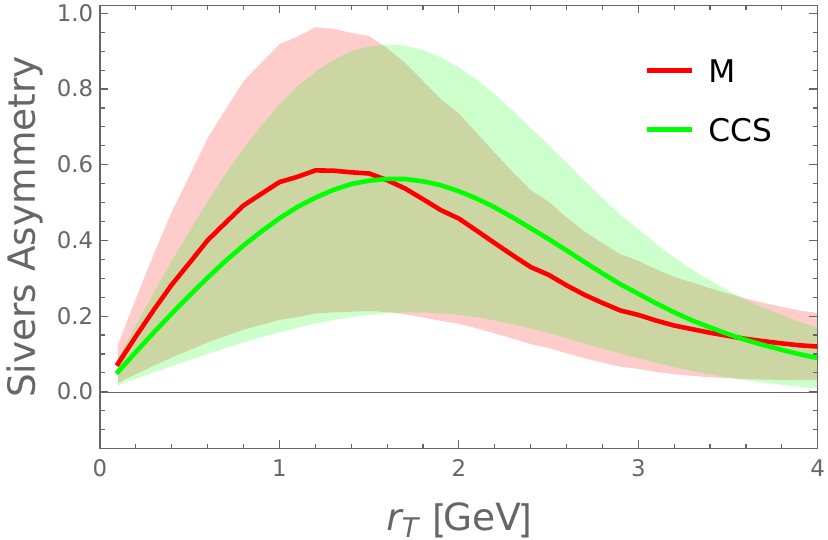}
    \caption{Sivers asymmetry with errors in the Sivers model. In CCS-scheme (green) and $\mathcal{M}$-scheme (red). 
    Hard scale errors is much smaller and not reported in the figure.}
    \label{fig:Xsecsaermod}
\end{figure}

%%%%%%%%%%%%%%%%%%%%%%%%%%%%%
The Sivers asymmetry is then shown in fig.~\ref{fig:Xsecsa}  with the two renormalization schemes and hard scale error bands. 
%%%%%%%%%%%%%%%%%%%%%%
%%%%%%%%%%%%%%%%%%%%%%
 Besides, we have estimated the Sivers model error using 500 replicas extracted from ref.~\cite{Bury:2021sue}, where for each replica they found the corresponding values of the non-perturbative parameters of the Sivers model. Using these non-perturbative parameters we obtained the polarized cross-section values and calculated the mean and standard deviation of it. As expected the errors are very big and the relative bands are shown in  fig.~\ref{fig:Xsecsaerr} with the two renormalization schemes.
%%%%%%%%%%%%%%%%%%%
%%%%%%%%%%%%%%%%%
The corresponding final results for the Sivers asymmetry is in fig.~\ref{fig:Xsecsaermod}.
The asymmetry can be large, having its peak for $r_T\sim$ 1-2 GeV with a value in the interval 5-96$\%$.
%%%%%%%%%%%%%%%%%%%%%
%%%%%%%%%%%%%%%%%%%%%

We have compared the predictions for the Sivers asymmetry obtained using the evolution kernels and distributions from the ART23~\citep{Moos:2023yfa} and ART25~\citep{Moos:2025sal} analyses.  The $\mathcal{M}$ scheme exhibits a stronger sensitivity to Collins-Soper kernel differences than the CCS scheme, however, given the current level of precision, the predictions of the two schemes remain compatible within uncertainties.

\begin{figure}[h]
    \centering
    \includegraphics[width=0.5\linewidth]{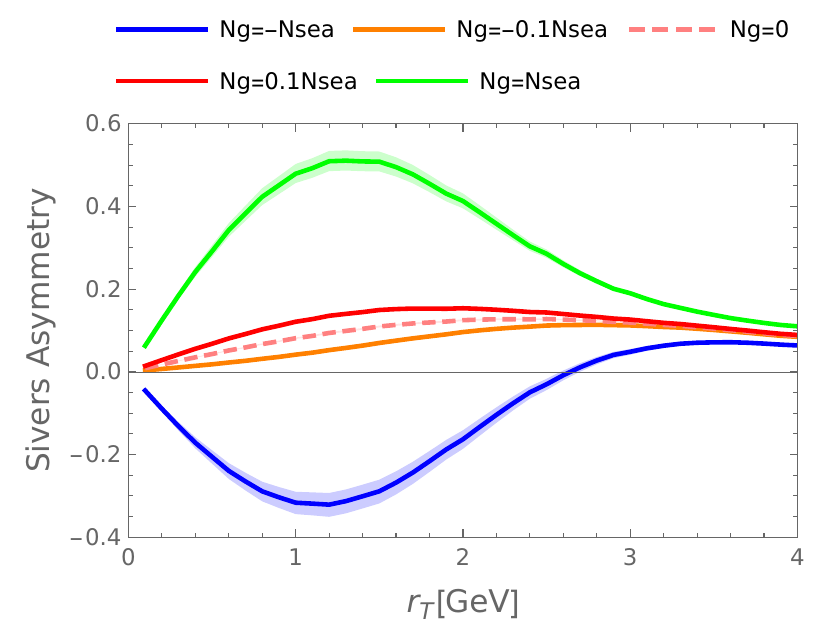}
    \caption{Sivers asymmetry with hard scale variation using the $\mathcal{M}$-scheme. The gluon-Sivers function has been scaled by different factors in each line.
    All lines have hard scale variation error bands.
    Model error is not reported.
    \label{fig:gmodels}}
\end{figure}

The results presented so far rely on the ansatz that the gluon Sivers distribution is equal to the sea-quark contribution extracted from fits of the quark Sivers functions. Varying the normalization of this model, including a possible sign change, leads to significantly different predictions, as illustrated in fig.~\ref{fig:gmodels}. In particular, setting the gluon Sivers contribution to zero (dashed line in fig.~\ref{fig:gmodels}) still yields an asymmetry at the level of a few percent. A similar magnitude is obtained when the sign of the gluon Sivers distribution is reversed relative to the default assumption (blue line in fig.~\ref{fig:gmodels}). An observed asymmetry below the percent level would therefore suggest a nontrivial cancellation between quark and gluon contributions.

%%%%%%%%%%%%%%%%%%%%%%%
\section{Conclusions}
\label{sec:Conclusions}
%%%%%%%%%%%%%%%%%
Dijet cross section in SIDIS is sensitive to gluon distributions. Here we have re-analyzed the unpolarized case and provided the polarized one  using TMD factorization at LP, and TMD evolution at N$^3$LO. We have identified several sources of uncertainty: scale dependence in hard and collinear-soft functions, non-perturbative  inputs of the TMD evolution kernels, scheme dependence of  evolution kernels, model dependence of the Sivers functions.
All of these items provide hints on the issues that future research must undertake to get a reliable estimations. This includes better perturbative control of hard factors and other functions appearing in the factorization theorem. 
The results that we obtain are still largely model dependent, especially on gluon distributions. As a guideline we have used sea-partonic distributions extracted from previous fits.
We are also proposing a resummation scheme (${\cal M}$-scheme) which is simpler than the one  adopted previously in the literature (CCS-scheme) and consistent with $\zeta$-prescription. The scheme is consistent with a SCET-II factorization, but does not resum sum parts which appear in SCET$_+$ scheme, basically the separation of collinear-soft and jet functions. The latter resummation introduces imaginary parts in the anomalous dimensions that require a sophisticated treatment and limitation on scale factorization intervals \citep{delCastillo:2021znl}. The ${\cal M}$-scheme provides in general smaller errors bands which suggest a better convergence and it is more sensitive to the precise value of the TMD evolution kernel.
This suggests that log resummation in collinear-soft and jet function factorization is not as relevant as figured out in  theoretical works. More  studies are necessary to arrive to more precise statements.
Despite all this, the main indication that we have is that Sivers asymmetry is expected to be large, above 5$\%$ which is clearly interesting phenomenologically and relevant for EIC.

\acknowledgments

We thank Rafael Fernández del Castillo for collaborating in the first stages of this work and Alexey Vladimirov for the useful discussions and support. 
P.A.G.G. is supported by the Ministry of Education contract FPI-PRE2020-094385.
This project is supported by the Spanish Ministerio de Ciencia, Innovación y Universidades through grants No. PID2022-136510NB-C31 and PID2022-136510NB-C33 funded by MCIN/AEI/10.13039/501100011033, as well as CNS2022-135186.
This project is supported as well by the European Union Horizon research MSCA–Staff Exchanges,
HORIZON-MSCA-2023-SE-01-101182937-HeI and 
COST action n. 24159 (SHARP), and by the Basque Government through the grant IT1628-22.

\appendix
\section{Appendix}
\subsection{$\zeta_{\mu}^{\mathcal{M}_i}$ solution in $\mathcal{M}$-scheme}
\label{app:Msadp}
We have the following equation for $\zeta_{\mu}^{\mathcal{M}_i}(b)$:
\begin{equation}\label{eq:zeta_mu_diff_eq}
    \gamma_{\mathcal{M}_{i}}(\mu, \zeta_{\mu}^{\mathcal{M}_i}(b)) = 2 \mathcal{D}_{\mathcal{M}_{i}} (\mu,b) \frac{d \ln \zeta_{\mu}^{\mathcal{M}_i}(b)}{d\ln \mu^2},
\end{equation}

To obtain the perturbative solution of $\zeta_{\mu}^{\mathcal{M}_{g,q}\text{pert}}(b)$ at first order in $\alpha_s$ we replace the value of $D_{\mathcal{M}_i}$ which is the same CS kernel as the TMD. We also replace the value of the anomalous dimension of the total function $\mathcal{M}$ for each channel given by eq.~(\ref{eq:GMs}) and obtain:
\begin{align}
    &\zeta_{\mu}^{\mathcal{M}_{g}\text{pert}}=\zeta_{2,0}^{\mathcal{M}_g} = \lp \frac{p_T^2 R^2}{\hat{s}} \rp^{2\frac{C_F}{C_A}} \,,\nonumber \\
    &\zeta_{\mu}^{\mathcal{M}_{q}\text{pert}}= \zeta_{2,0}^{\mathcal{M}_q} = \lp \frac{p_T^2 R^2 }{\hat{s}} \rp^{\frac{C_F + C_A}{C_F}} \lp \frac{\hat{t}}{\hat{u}} \rp^{\frac{C_F - C_A}{C_F}} \,,
\end{align}
and there is no dependence on the scale $\mu$. 
To obtain the exact solution we follow the same procedure as in~\cite{Vladimirov:2019bfa}, using  the change to variable $g= \ln\frac{\zeta}{\zeta_0}$, 
at the current order (\ref{eq:zeta_mu_diff_eq}) for the gluon case yields:
\begin{equation}
    -4 C_A g_0 = 2 \mathcal{D}_{\mathcal{M}_{i}}\lb -\beta_0 g_0 -\frac{\Gamma_0}{2}g'_0\rb \,.
\end{equation}
 We have so the solution $g_0=0$ which brings to
\begin{align}
    &\zeta_{\mu}^{\mathcal{M}_g,\text{exact}}=\zeta_{\mu}^{\mathcal{M}_g,\text{pert}} \,, 
    &\zeta_{\mu}^{\mathcal{M}_q,\text{exact}}=\zeta_{\mu}^{\mathcal{M}_q,\text{pert}}.
\end{align}
The non null  nontrivial  solution however does not meet the border conditions that $g=0$ when $\zeta=\zeta_0$.

\subsection{Angular integrals }
\label{app:angularintegrals}
We express the angular integrals of 
 eq.~(\ref{eq:M1g}-\ref{eq:M1q}) as
\begin{itemize}
    \item \textit{Single logarithmic terms},
    \begin{equation}
       I_{log} =\int_{-\pi}^{\pi} d\phi_b \sin^2(\phi_S - \phi_b) \ln(-i\cos\phi_b)= -\frac{\pi}{2} \left(\cos(2\phi_S) + \ln(4)\right) \,,
    \end{equation}
    where,
    \begin{equation}
        \ln(-i\cos\phi_b) = \ln|\cos\phi_b|-\frac{i\pi}{2} \Theta(\phi_b)\,.
    \end{equation}
    \item \textit{Double logarithmic terms}:
    \begin{equation}
         I_{log2} = \int_{-\pi}^{\pi} d\phi_b \sin^2(\phi_S - \phi_b) \ln^2(-i\cos\phi_b)= -\frac{\pi^3}{6} + \pi \ln^2 2 + \frac{1}{2} \pi \cos(2 \phi_S) (1 + \ln4)\,.
    \end{equation}
    \item \textit{Single Logarithmic $A_b$ terms},
    \begin{align}
        I_{A_b}    &=\int_{-\pi}^{\pi} d\phi_b \sin^2(\phi_S - \phi_b) \ln(-A_b) \nonumber \\
         &= -\cos 2\phi_s \Big[ \ln\left(\frac{\hat{s}}{4p_T^2} \right) I_0(1) - 2 I_1(1)  \Big] + \cos^2\phi_S \Big[\ln\left(\frac{\hat{s}}{4p_T^2} \right)  I_0(0) -  2I_1(0)\Big]\,.
    \end{align}
    \item \textit{Double Logarithmic $A_b$ terms},
    \begin{align}
       I_{A_b^2}=  &\int_{-\pi}^{\pi} d\phi_b \sin^2(\phi_S - \phi_b) \ln^2(-A_b) = \pi\ln^2\left(\frac{\hat{s}}{4 p_T^2} \right) \nonumber \\
         &+ 4\ln\left(\frac{\hat{s}}{4p_T^2}\right) \Big(\cos(2\phi_S)I_1(1) - \cos^2\phi_S I_1(0)\Big) + 4 [I_2(0) \cos^2\phi_S -I_2(1)\cos2\phi_S ]\,,
    \end{align}
    with
     \begin{equation}
    A_b = \frac{ (v_1 \cdot v_2)}{2\, ( v_1 \cdot \hat{b} ) ( v_2 \cdot \hat{b})} = -\frac{\hat{s}}{4 p_T^2 (\cos{\phi_b})^2}\,.
    \end{equation}
    \item \textit{Poly-logarithmic terms},
    \begin{align}
      I_{Li2} &=\int_{-\pi}^{\pi} d\phi_b \sin^2(\phi_S - \phi_b) \text{Li}_2\left(1-\frac{\hat{s}}{4p_T^2 \cos^2\phi_b}\right) \nonumber \\ &
      = 
      \int_{-\pi}^{\pi}d\phi_b \left(-\cos 2\phi_S\cos^2\phi_b +\cos^2\phi_S \right)  \text{Li}_2\left(1-\frac{1}{\bar{c} \cos^2\phi_b}\right) \nonumber\\
      &= -\cos 2\phi_S I_{Li}(1,\bar{c}) + \cos^2\phi_SI_{Li}(0,\bar{c})\,,
    \end{align}
\end{itemize}
where
\begin{align}
    \bar{c} &\equiv \frac{4p_T^2}{\hat{s}}\;,& 0&\,<\,\bar{c}\,\leq\,1,
\end{align}
and
\begin{equation}
    I_{\text{Li}}(\mathcal{A}, \bar{c})\equiv   \int_{-\pi}^{+\pi} d\phi_b\,  |\cos\phi_b|^{2 \mathcal{A}} \; \text{Li}_2 \lp1 - \frac{1}{\bar{c}\; \cos^2\phi_b}\rp,
\end{equation}
which has been solved in \cite{delCastillo:2021znl} and is equal to
\begin{multline}
     I_{\text{Li}}(\mathcal{A}, \bar{c} ) =  -\frac{1}{2} \lb \lp \ln^2\bar{c} +\frac{\pi^2}{3}  \rp I_0(\mathcal{A}) + 4 I_1(\mathcal{A})\, \ln \bar{c}  + 4 I_2(\mathcal{A})  \rb  \\[5pt]
     -\sum_{n=1}^{\infty} \frac{{\bar{c}}^{\,n}}{n} \lb \lp \ln \bar{c} -\frac{1}{n} \rp I_0(\mathcal{A}+n) + 2 I_1(\mathcal{A}+n)\rb  .
\end{multline}
We have also used,
\begin{equation}
    \Theta(\phi_b) = 
    \begin{cases}
    +1  &:  -\pi/2 < \phi_b < \pi/2 \,,\\
    -1 &: \text{otherwise} \,.
    \end{cases}
\end{equation}
We also use the following basic integrals:
\begin{equation}
  I_n (\mathcal{A})\equiv  \int_{-\pi}^{+\pi} d\phi_b\,  |\cos\phi_b|^{2 \mathcal{A}} \; \ln^{n} |\cos\phi_b| \,,
\end{equation}
where 
\begin{align}\label{eq:integral_basis}
     I_0(\mathcal{A}) &= \frac{2 \sqrt{\pi} \;\Gamma (1/2+\mathcal{A} )}{\Gamma(1+\mathcal{A})},   \nl
     I_1(\mathcal{A}) &= \frac{\sqrt{\pi} \;\Gamma(1/2+\mathcal{A})}{\Gamma(1+\mathcal{A})} \; (H_{\mathcal{A}-1/2} - H_{\mathcal{A}}) \nl
     I_2(\mathcal{A}) &= \frac{\sqrt{\pi} \;\Gamma(1/2+\mathcal{A})}{2 \Gamma(1+\mathcal{A})} \; \lb (H_{\mathcal{A}-1/2} - H_{\mathcal{A}})^2 +\psi^{(1)}\lp\frac{1}{2} + \mathcal{A}\rp - \psi^{(1)}(1+\mathcal{A}) \rb  ,
\end{align}
where $\mathcal{A} > -1/2$, $H$ is the harmonic function and $\psi$ is the polygamma function
\begin{equation}
    \psi^{(n)}z) = \frac{d^n}{dz^n} \psi(z) = \frac{d^{n+1}}{dz^{n+1}} \ln \Gamma(z)
\end{equation}

\section{Hard prefactors}
\label{app:Prefactors}
The hard prefactors for each channel are given in ref.~\cite{Pisano:2013cya,Boer:2016fqd,Efremov:2017iwh,Efremov:2018myn},
\begin{align}
    \sigma_0^{g U} = 2 \pi p_T \frac{\mathcal{N}}{x s}A_0^{gU},\qquad \sigma_0^{q U}=2 \pi p_T \frac{\mathcal{N}}{x s} A_0^{qU} ,\qquad     \sigma_0^{g L} = -4\pi p_T \frac{\mathcal{N}}{x s} B_2, 
\end{align}
where
\begin{equation}
    \mathcal{N}=\frac{\alpha^{2} \alpha_{s}}{\pi s p_T^{2}} \frac{1}{x y^{2}},
\end{equation}
\begin{align}
    A_0^{gU} & = e_q^2 T_R \lb \lp 1 + (1-y^2) \rp A_{U+L}^{gU} - y^2 A_{L}^{gU} \rb , \\
    A_0^{qU} & = e_q^2 C_F \lb \lp 1 + (1-y^2) \rp A_{U+L}^{qU} - y^2 A_{L}^{qU} \rb ,\\
    B_{2}&= e_{q}^{2} T_{R} \lb \lp 1 + (1-y^2) \rp B_{U+L} - y^2 B_{L} \rb \frac{r_T^2}{M_{p}^{2}},
\end{align}
and $A$ and $B$ factors are given by
\begin{align}
    A_{U+L}^{qU}&=\frac{1-z}{D^{2}}\left\{1+z^{2}+\left[2 z(1-z)+4 z^{2}(1-z)^{2}\right] \frac{Q^{2}}{p_T^2}+\left[z^{2}(1-z)^{2}\right]\left[1+(1-z)^{2}\right] \frac{Q^{4}}{p_T^4}\right\},\\
    A_{U+L}^{gU}&=\frac{1}{D^{3}}-\frac{z(1-z)}{D^{3}}\left\{2-8 z(1-z) \frac{Q^{2}}{p_T^{2}}\right.
    \left.-z(1-z)[1-2 z(1-z)] \frac{Q^{4}}{p_T^{4}}\right\},\\
    B_{U+L}&=\frac{z(1-z)}{D^{3}}\left\{[1-6 z(1-z)] \frac{Q^{2}}{p_T^{2}}\right\},\\
    A_{L}^{qU}&=4 \frac{z^{2}(1-z)^{3}}{D^{2}} \frac{Q^{2}}{p_T^{2}},\\
    A_{L}^{gU}&=8 \frac{z^{2}(1-z)^{2}}{D^{3}} \frac{Q^{2}}{p_T^{2}},\\
    B_{L}&=-4 \frac{z^{2}(1-z)^{2}}{D^{3}} \frac{Q^{2}}{p_T^{2}},
\end{align}
where $D$ is defined as
\begin{equation}
D=1+z(1-z) \frac{Q^{2}}{p_T^{2}}.
\end{equation}
In order to pass from the expressions of cross section of these works to ours we integrate in inelasticity using
\begin{align*}
    \int dy\; \delta(y x s-Q^2).
\end{align*}

%%%%%%%%%%%%%%%%%%%%%%%%
%%%%%%%%%%%%%%%%%%%%%%
\section{Anomalous dimensions}
\label{app:AD}
%%%%%%%%%%%%%%%%%%%%%%%%
The one loop anomalous dimensions used in this paper are ($f=q,\;\bar q$)
\begin{align} 
\gamma_{H_{\gamma g} }^{[1]} &= 4 \lbc   C_F \lb \ln \lp  \frac{ \hat{s} ^2}{\mu^4} \rp  -2 \gamma_q   \rb  + C_A  \ln \lp \frac{\hat{t} \,\hat{u} } {\hat{s} \mu^2} \rp  \rbc \,, \nl
\gamma_{H_{\gamma f} }^{[1]}  & = 4 \lbc  C_F \lb \ln\lp  \frac{\hat{u}^2}{\mu^4}\rp  - 2 \gamma_q  \rb  + C_A   \ln \lp \frac{\hat{s} \, \hat{t}}{\hat{u}\, \mu^2} \rp   \rbc \,,\nl
\gamma_{S_{\gamma g}} ^{[1]}  &= 4 \lbc  - C_A \ln \zeta_2  +  2 C_F \lb \ln (B \mu^2 \,e^{2\gamma_E})  +\ln \frac{p_T^2}{\hat s} +\ln(4 c_{\bmat{b}}^2 )\rb \rbc\,, \nl
\gamma_{S_{\gamma f}} ^{[1]}   & =  4 \lbc  (C_F +C_A) \lb  \ln (B  \mu^2 e^{2\gamma_E})  +\ln \frac{p_T^2}{\hat s}  + \ln (4 c_{\bmat{b}}^2 ) \rb  + (C_F-C_A)\lb  \ln \lp \frac{\hat{t} }{ \hat{u}  }  \rp   -  \kappa(v_f) \rb- C_F \ln\zeta_2    \rbc  \,, \nl
\gamma_{F_i}^{[1]}       &=4 C_i \lb - \ln\lp \frac{\zeta_1}{\mu^2} \rp + \gamma_i \rb \,,\nl
\gamma_{J_i}^{[1]}   &= 4 C_i \lb   -\ln \lp \frac{p_T^2}{\mu^2} \rp  -\ln R^2 + \gamma_i  \rb\,, \nl
\gamma_{\mathcal{C}_g}^{[1]}&= 4 C_A \lb   -  \ln \lp B \mu^2 \,e^{2\gamma_E} \rp  + \ln R^2 -\ln (4 c_{\bmat{b}}^2 )  + \kappa(v_g)\rb \,,\nl
\gamma_{\mathcal{C}_i}^{[1]}&= 4 C_F \lb   -  \ln \lp B \mu^2 \,e^{2\gamma_E} \rp  + \ln R^2 -\ln (4 c_{\bmat{b}}^2 )  + \kappa(v_i)\rb \,,\nonumber \\[8pt]
\gamma_\alpha^{[1]} &= - 4 C_A \gamma_g\,,
\end{align} 
The imaginary component in the soft and collinear-soft anomalous dimension is denoted by $\kappa(v_i)$ where 
\begin{align}
\kappa(v_f) = - \kappa(v_{\bar{f}}) = - \kappa(v_g) = i \pi\, \text{sign}( c_{\bmat{b}}).
\end{align}
These anomalous dimensions can be found in \cite{Becher:2009th,Becher:2012xr,Chien:2020hzh,Hornig:2016ahz,Buffing:2018ggv,Echevarria:2015byo,delCastillo:2020omr,delCastillo:2021znl}.
\section{Sivers fit parameters}
\label{app:bury}
In our results the default values are taken from the  DY+SIDIS fit in \cite{Bury:2020vhj} at N$^3$LO and reported in tab~\ref{tab:bury}. 

\begin{table}[htb]
\renewcommand{\arraystretch}{1.2}
\begin{center}
\begin{tabular}{l||c|c||c|c|}
\multirow{2}{*}{Parameter} & SIDIS & DY+SIDIS & SIDIS &DY+SIDIS
\\ & at NNLO & at NNLO & at N$^3$LO & at N$^3$LO
\\\hline
$r_0$  	& $0.95_{-0.94}^{+0.70}$ 	& $0.58_{-0.57}^{+0.71}$ 	& $0.94_{-0.93}^{+0.71}$ 	& $0.54_{-0.53}^{+0.60}$
\\\hline
$r_1$ 	& $0.09_{-0.09}^{+5.90}$  	& $4.8_{-3.1}^{+1.9}$ 		& $1.02_{-1.02}^{+4.96}$	& $5.22_{-3.43}^{+1.18}$
\\\hline	
$r_2$ 	& $195._{-20.}^{+434.}$		& $192._{-119.}^{+101.}$ 		& $223._{-47.}^{+409.}$		& $203._{-133.}^{+71.}$
\\\hline
$N_u$ 	& $-0.013_{-0.007}^{+0.008}$& $-0.020_{-0.020}^{+0.013}$ & $-0.012_{-0.008}^{+0.007}$ &$-0.017_{-0.023}^{+0.011}$
\\\hline
$\beta_u$ & $-0.35_{-0.06}^{+0.07}$	& $-0.35_{-0.10}^{+0.09}$ 	& $-0.33_{-0.07}^{+0.06}$	& $-0.36_{-0.11}^{+0.09}$
\\\hline
$\epsilon_u$ & $-3.9_{-0.3}^{+0.5}$	& $-3.9_{-0.6}^{+0.6}$ 		& $-3.8_{-0.4}^{+0.4}$		& $-3.9_{-0.6}^{+0.6}$
\\\hline
$N_d$ 	& $0.34_{-0.21}^{+0.25}$	& $0.40_{-0.18}^{+0.20}$ 	& $0.34_{-0.21}^{+0.25}$	& $0.37_{-0.17}^{+0.18}$
\\\hline
$\beta_d$& $-0.77_{-0.13}^{+0.61}$	& $-0.51_{-0.29}^{+0.70}$ 	& $-0.82_{-0.08}^{+0.66}$	& $-0.7_{-0.11}^{+0.77}$
\\\hline
$\epsilon_d$& $1.8_{-2.8}^{+20.0}$	& $9.4_{-9.9}^{+13.9}$  	& $3.6_{-4.7}^{+18.4}$		& $9.0_{-9.1}^{+17.6}$
\\\hline
$N_s$ 	& $0.43_{-0.34}^{+0.42}$	& $0.90_{-0.56}^{+0.83}$ 	& $0.48_{-0.38}^{+0.37}$	& $0.76_{-0.43}^{+0.89}$
\\\hline
$N_{\text{sea}}$& $-0.23_{-0.21}^{+0.15}$& $-0.51_{-0.31}^{+0.22}$ & $-0.23_{-0.22}^{+0.15}$& $-0.47_{-0.32}^{+0.21}$
\\\hline
$\beta_{s}=\beta_{\text{sea}}$& $2.3_{-1.2}^{+0.7}$& $2.5_{-0.7}^{+0.8}$ & $2.3_{-1.2}^{+0.7}$& $2.5_{-0.7}^{+0.8}$
\end{tabular}
\end{center}
\caption{\label{tab:bury} Values of parameters obtained in various fits of ref.~\cite{Bury:2020vhj}. }
\end{table}
\begin{table}[htb!]
\def\arraystretch{1.25}
  \begin{center}
    \begin{tabular}{c |c c || c |c|}
    %    & \multicolumn{3}{c}{$\chi^2/d.o.f.= 1.032$} & \\       \hline
        $N_{u}$         & $ 0.077_{-0.005}^{+0.004}$ GeV & & $\alpha_{u}$     &\hspace{-1.2cm} $ 0.967_{-0.045}^{+0.028}$\\
        $N_{d}$         & $-0.152_{-0.016}^{+0.017}$ GeV& & $\alpha_{d}$     & \hspace{-1.2cm}$ 1.188_{-0.023}^{+0.056}$\\
        $N_{s}$          & $ 0.167_{-0.051}^{+0.053}$ GeV& & $\alpha_{sea}$ & $ \hspace{-1.2cm}0.936_{-0.026}^{+0.069}$\\
        $N_{\bar{u}}$  & $-0.033_{-0.017}^{+0.016}$ GeV& & $\beta$       & \hspace{-1.2cm} $5.129_{-0.034}^{+0.017}$\\
        $N_{\bar{d}}$ & $-0.069_{-0.026}^{+0.019}$ GeV& & $g_1^T\,$ &  $0.180_{-0.070}^{+0.035}$ GeV$^2$\\
        $N_{\bar{s}}$ & $-0.002_{-0.040}^{+0.047}$  GeV& & 
    \end{tabular}
    \caption{Fit parameters for ref.~\cite{Echevarria:2020hpy}.}
    \label{tab:echevarria}
  \end{center}
\end{table}
\section{Model EKT
}
\label{sec:ekt}
 
The model  of ref.~\cite{Echevarria:2020hpy} is obtained performing first
an operator product expansion of  the Sivers function onto the Qiu-Sterman function it includes a non-perturbative $b_T$-dependent correction,
\begin{align}
    f_{1T,\text{EKT}}^{\perp} & (\xi, b;\mu,\zeta) = \Big[\bar{C}_{q\leftarrow i} \otimes T_{F\, i/p}\Big](\xi,b;\mu,\zeta) e^{-g_1^T b^2} \,,
    \label{e.Sivers}
\end{align}
where $\bar{C}$ is a Wilson coefficient, perturbatively calculable, and $T$ the non-perturbative Qiu-Sterman function.

The convolution involves two kinematic variables $\hat{x}_1$ and $\hat{x}_2$ and it is given by
\begin{align}
&\Big[\bar{C}_{q\leftarrow i} \otimes T_{F\, i/p}\Big](x,b;\mu,\zeta) = 
\int_{x}^{1} \frac{d\hat{x}_1}{\hat{x}_1}\frac{d \hat{x}_2}{\hat{x}_2} \bar{C}_{q\leftarrow i}(x/\hat{x}_1,x/\hat{x}_2,b;\m,\z) \, 
T_{F\, i/p}(\hat{x}_1,\hat{x}_2;\m)
\,.
\end{align}
For $\mu^2 = \mu_{b_*}^2$
the coefficient $\bar{C}$ is 
\begin{align}
\bar{C}_{q\leftarrow q'}(x_1,x_2,b;\mu_{b_*},\mu_{b_*}^2)  = & \delta_{qq'} \delta(1-x_1) \delta(1-x_2)-\frac{\alpha_s}{2\pi} \frac{\delta_{qq'}}{2N_C}\d(1-x_2/x_1)(1-x_1)
\,\nn \\
& -\frac{\alpha_s}{2\pi}\delta_{qq'} C_F \frac{\pi^2}{12}\d(1-x_1)\d(1-x_2).
\end{align}
and 
the result of the convolution is:
\begin{align}
\Big[\bar{C}_{q\leftarrow i} \otimes T_{F\, i/p}\Big](x,b;\mu,\zeta) &= \delta_{qi} \lp 1-C_F\frac{\alpha_s}{2\pi}  \frac{\pi^2}{12}\rp T_{F\, i/p}({x},{x};\m) \nn
\\ &
- \frac{\alpha_s}{2\pi}\frac{\delta_{qi}}{2N_C}
\int_{x}^{1} \frac{d\hat{x}_1}{\hat{x}_1} \lp 1 - \frac{x}{\hat{x}_1}\rp\, 
T_{F\, i/p}(\hat{x}_1,\hat{x}_1;\m)
\,.
\end{align}

Furthermore, $T_{F\, q/p}(x, x, \mu)$ is assumed to be proportional to the unpolarized PDF $f_{q/p}(x,\mu_0)$,
\begin{align}
    T_{F\, q/p}(x,x,\mu) = \mathcal{N}_q(x)f_{q/p}(x,\mu) \,,
\end{align}
with $\mathcal{N}_q(x)$ given by 
\begin{align}
    \mathcal{N}_q(x) = N_q\frac{\left( \alpha_q+\beta_q \right)^{\left( \alpha_q+\beta_q \right)}}{\alpha_q^{\alpha_q} \beta_q^{\beta_q}}x^{\alpha_q}(1-x)^{\beta_q}\,.
\end{align}
The value of the parameters are reported in  tab.~\ref{tab:echevarria}.

%%%%%%%%%%%%%%%%%%%%%%%%%%%%%%%%%%%%%%%%%%%%%%%%%%%%%%%%%%%%%%%%%%%%%%%%%%%%%%%%%%
\bibliographystyle{JHEP}
\normalbaselines 
\bibliography{TMD_ref}
%%%%%%%%%%%%%%%%%%%%%%%%%%%%%%%%%%%%%%%%%%%%%%%%%%%%%%%%%%%%%%%%%%%%%%%%%%%%%%%%%%

\end{document}